\documentclass[twocolumn,pre,aps,nofootinbib]{revtex4}
\usepackage{amsmath}
\usepackage[final]{graphicx}
\usepackage{amsfonts}

 \let\b=\beta \let\g=\gamma \let\d=\delta
 \let\z=\zeta \let\h=\eta 
\let\l=\lambda    
 \let\t=\tau \let\f=\varphi 
   \let\G=\Gamma
\let\D=\Delta   
   
   \let\io=\infty

\def\ie{{i.e. }}

 \def\xx{{\bf x}}

\def\to{\rightarrow} \def\la{\left\langle} \def\ra{\right\rangle}

\newcommand{\beq}{\begin{equation}} \newcommand{\eeq}{\end{equation}}
 \newcommand{\wt}{\widetilde}

\begin{document}

\title{
Dynamical modelling of molecular constructions and setups for DNA unzipping}

\author{Carlo Barbieri}
\author{Simona Cocco}
\affiliation{LPSENS, Unit\'e Mixte de Recherche (UMR 8550) du CNRS et
  de l'ENS, associ\'ee \`a l'UPMC Universit\'e Paris 06, 24 Rue Lhomond, 75231
  Paris Cedex 05, France.}

\author{R\'emi Monasson}
\author{Francesco Zamponi}
\affiliation{LPTENS, Unit\'e Mixte de Recherche (UMR 8549) du CNRS et
  de l'ENS, associ\'ee \`a l'UPMC Universit\'e Paris 06, 24 Rue Lhomond, 75231
  Paris Cedex 05, France.}

\begin{abstract}
We present a dynamical model of DNA mechanical unzipping under the
action of a force. The model includes the motion of the fork in the sequence-dependent
landscape, the trap(s) acting on the bead(s), and the polymeric components of the
molecular construction (unzipped single strands of DNA, and linkers). Different
setups are considered to test the model, and the outcome of the simulations is
compared to simpler dynamical models existing in the literature where polymers
are assumed to be at equilibrium.
\end{abstract}

\maketitle

\section{Introduction}

\begin{figure}[htbp]
\includegraphics[width=8cm]{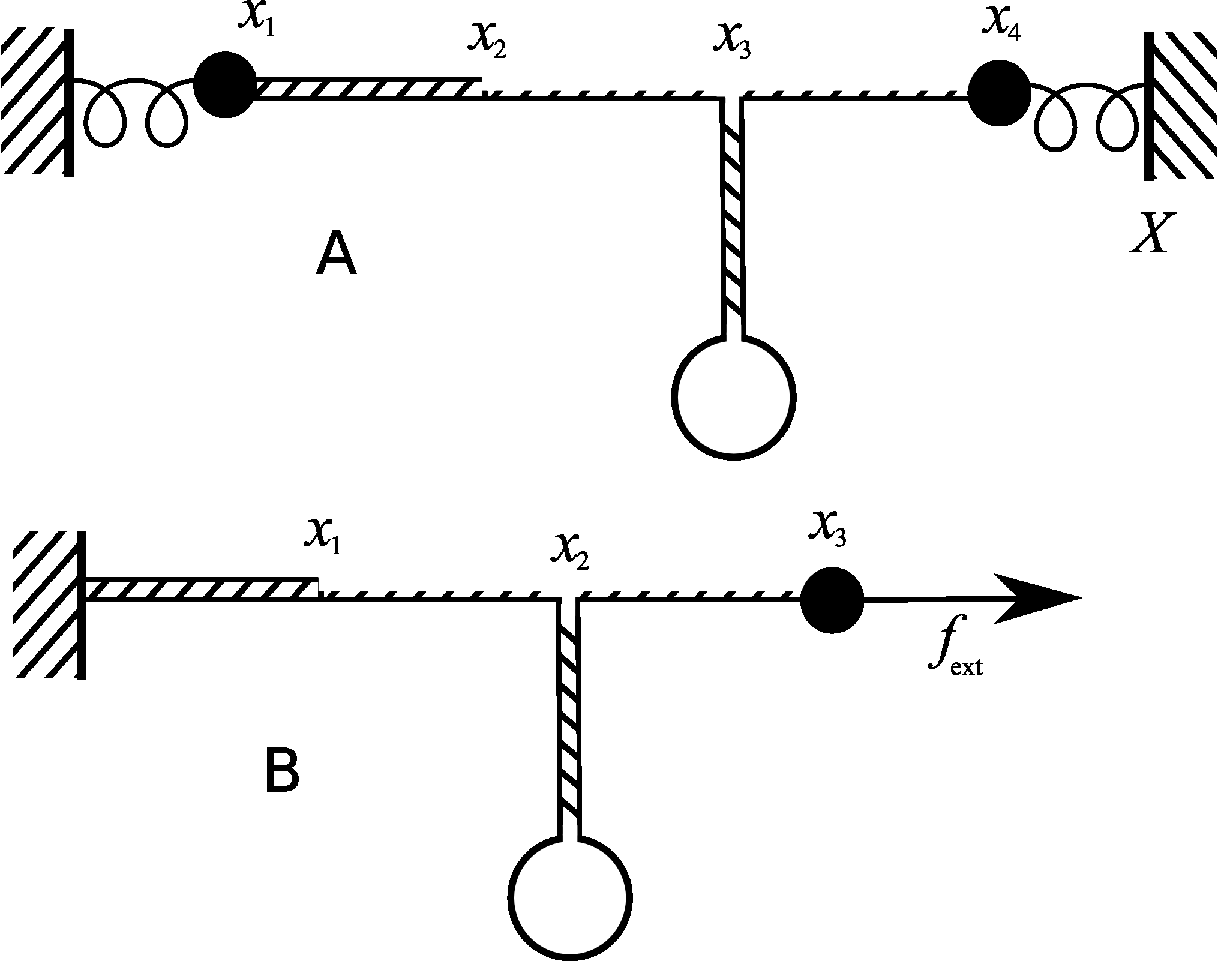}
\caption{Typical experimental setups that will be described in
the following. A) A setup with two optical traps (beads $x_1$ and
$x_4$) drawn as springs and whose centers are the black vertical lines;
B) a setup with a single magnetic bead $x_3$ that applies
a constant force on the molecule attached to a fixed ``wall''. 
In both cases the molecular
construction is made by a DNA molecule that has to be opened
(therefore one should include two single-strand linkers that are
the opened parts of the molecule) and one double-stranded DNA linker.
The coordinates $x_i$ are the distances of the corresponding points
from the left reference position (which is the center of the left optical
trap in case A and the fixed wall in case B).
}
\label{fig:setup}
\end{figure}

Over the past fifteen years various single molecule experiments have 
investigated DNA mechanical and structural properties
~\cite{mb,mb2,Bus03,Coc21,Smi92,Clu96,Smi96,Ess97,Boc98,Boc02,Tho02,Boc04,Man06,Man07,Lip01,Dan03,Dan04,Wee05}, and protein-DNA interactions
~\cite{Har03,Van03,Per03,Wui00,Mai00,Lev03,Lan03,Sau03,Mat04,Lio06,Lio07}.
These experiments provide dynamical information usually hidden in
large scale bulk experiments, such as fluctuations on the scale of the
individual molecule. The separation of the two strands of a DNA
molecule under a mechanical stress, usually referred to as unzipping,  
has first been carried out by Bockelmann and Heslot in 1997 \cite{Ess97}.  The
strands are pulled apart at a constant velocity while the force
necessary to the opening is measured.  The average opening force for
the $\lambda$ phage sequence is of about 15 pN (at room temperature and standard ionic conditions), with fluctuations 
around this value that depend on the particular sequence
content. Bockelmann, Heslot and collaborators have shown that the
force signal is correlated to the average sequence on the scale of ten
base-pairs but could be affected by the mutation of one base-pair
adequately located along the sequence \cite{Boc02}.  Liphart et
al. and \cite{Lip01} and Danilowicz et al. \cite{Dan03,Dan04,Wee05} have performed
an analogous experiment, using a constant force setup, on a short RNA
and a long DNA molecules respectively (Figure \ref{fig:setup}B). The
distance between the two strands extremities is measured as a function
of the time while the molecule is submitted to a constant force.  DNA
strands separation has been also studied in single molecule
experiments by translocation through nanopores \cite{Sau03,Mat04}. 

The motivation underlying unzipping experiments of DNA is (at least)
two-fold. First the study of unzipping aims at a better understanding 
of the mechanisms governing the opening of
DNA during transcription and replication by proteins such as
Polymerases, Helicases, and Exonucleases
\cite{Van03,Per03,Lio06,Lio07}.  
Simple theoretical models describing the opening as
an unidimensional random walk on a sequence-dependent free energy
landscape have been proved to describe quite well several experimental
effect such as stick-slip motion in the opening at constant velocity
\cite{Boc98,Boc02}, the long pauses at fixed position of unzipping at
constant forces \cite{Lub,Coc4,Dan03}, the hopping dynamics between two
or more states in unzipping at critical forces of short DNA
molecules \cite{Lip01,Coc4,Mang07,Gre05}, and torsional drag effects in
unzipping at large velocity \cite{Tho02,Coc02}.  Moreover 
statistical mechanical
analysis have been successfully applied to extract from
experimental data the sequence-dependent free energy landscape and the
height of free energy barriers \cite{Col05,Woo06}.  

Secondly unzipping experiments could potentially be useful to extract
information on the sequence itself \cite{Bal06}.  Recently single
molecule sequencing has been achieved by monitoring a DNA/RNA
polymerase in the course of DNA synthesis from a ssDNA template
\cite{Har08,Gre05}; such single molecule sequencing could become
competitive with standard DNA sequencing because they do not require,
{\em a priori}, amplification through polymerase chain reactions.  A
fundamental question on the possibility of extracting information on
the sequence from unzipping experiments is the influence of the
experimental setup on the measures and the limitations imposed by the
latter \cite{siggia,Bal06}.  Indeed characteristic spatio-temporal
limitations are the finite rates of data acquisition, the relaxation
time of the bead, the limited spatial resolution, the thermal drift,
and more generally the noise in the instruments.
Moreover the dynamics of the opening fork (Figure \ref{fig:setup}) is influenced
by the single strands (open parts) of the molecule and the linkers, and
cannot be deduced directly from the observation of the bead from which 
the force or the position are measured.  

The accuracy of unzipping experiments at fixed velocity have improved
a lot over the last decade. Initially performed with an optical fiber 
\cite{Ess97}, experiments were then based on the use of simple optical
traps \cite{Boc02}. Nowadays double optical traps \cite{Woo06,Man06}
allow to reduce considerably the drift of the setup and to achieve a
temporal resolution of the order of 10 KHz, a sub-nanometric 
spatial resolution, and a precision on measured forces of the
order of fraction of pN.  Unzipping at fixed force has been
performed by a magnetic trap with a low temporal resolution (from 60
Hz to 200 Hz) due to the time needed to extract the position of the
bead, the spatial precision being of the order of
10 nm/$\sqrt{\mbox{Hz}}$) \cite{Lio06,Lio07}, or by an optical trap
also with a low temporal resolution (about 10 Hz) imposed by a feedback
mechanism needed to keep the force constant \cite{Lip01}. Recently a
new dumbbell dual optical trap has been developed. It operates without
feedback and can maintain the force constant over distances of about 50 nm
\cite{Gre05} with temporal resolution of 10 kHz and a spatial
resolution of 0.1 nm/$\sqrt{\mbox {Hz}}$.  

Limitations due to the
experimental systems were first addressed in \cite{siggia}. This
paper stated the impossibility to infer the sequence 
due to ssDNA fluctuations:  
fluctuations increase with the number of opened base-pairs and can become larger than the length of about 1 nm
corresponding to the spatial resolution of one open base-pair.  
This problem could however be solved by integrating out the single strand
dynamical fluctuations.  Several works have studied the effects of the setup
on the hopping dynamics of small RNA molecules
\cite{siggia,Gre05,Mang07,Hye08}.  The following effects have been
underlined. First of all the free energy landscape changes when adding to the free
energy an harmonic potential, due to the bead and handles,
\cite{Boc02,Mang07,Gre05,Hye08}. Therefore, for a given force, the measured
separation of the extremities depends on the stiffnesses of the trap and
handles.
Moreover, the opening and closing rates depend on the stiffness of the
optical trap; in particular when the experimental system gets softer
the fluctuations of the force gets smaller, and the hopping rates
approach their fixed-force values.

In this paper we introduce a model for the
coupled dynamics of the opening fork, the ssDNA strand, the
linkers, and the bead in the optical or magnetic trap. Essential notions and
existing literature are reviewed in Section II. Our dynamical model is presented
in Section III. Our program allows to simulate a generic setup, characterized by bead
dimensions, optical stiffness (absent in case of magnetical tweezers),
linker composition (dsDNA or ssDNA) and lengths, length of molecule to
be unzipped. All the parameters that characterize the
different dynamical components can be adjusted in the simulation.
The model is then used to simulate fixed-force (Section IV) and fixed-extension 
(Section V) numerical unzippings.

\section{Free-energies, time scales, and effective dynamics}

We discuss hereafter the thermodynamic properties of the various parts
of the experimental set-up (DNA sequence, open part of the molecule,
single- or double-strand linkers), as well as the relevant time
scales. Finally we briefly review previous dynamical studies where the
linkers and the open portion of the molecules are assumed to be at
equilibrium.


\subsection{Thermodynamics of the components}\label{thermo}

\subsubsection{Polymeric models for the linkers and open molecule}
\label{poly-models}

A polymer model is specified by its free energy as a function of the
extension $x$ for a given number $n$ of monomers; we call this 
quantity $W(x,n)$. When $x$ and $n$ are large $W$ is an extensive 
quantity hence $W(x,n) = n w(x/n) = n w(l)$, where 
$l=x/n$ is the extension per monomer. 
We also define
\beq\label{formulas-poly}\begin{split}
& f(l) = \frac{\partial W(x,n)}{\partial x} = w'(l) \ , \\
& l(f) = \text{ inverse of } f(l) \ , \\
& g(f) = \max_l [ f \, l - w(l) ] = f \, l(f) - w[l(f)] 
\end{split}\eeq
which are, respectively, the force at fixed extension, the average 
extension at fixed force, and the free-energy at fixed force. Notice that 
$g(f)$ is simply the integral of $l(f)$. Hence a polymer model is
completely described from the knowledge of the extension versus force
characteristic curve, $l(f)$.
In the following we will use some classical models for this function:
\begin{itemize}
\item{\it Gaussian (Hook) model:} 
\beq
l_{Hook}(f) = \frac f {k^m} \ .
\eeq
where the stiffness constant $k^m$ is related to the temperature $T$ and the 
average squared monomer length (at zero force) $b^2$ through $k^m = k_BT/b^2$. 
\item{\it Freely-Jointed Chain (FJC) model:}
\beq
l_{FJC}(f) =  \coth \left( \frac{f b}{k_BT} \right) -  \frac{k_BT}{f b} 
\eeq 
is the extension (per monomer) of a chain of rigid rods of length $b$, 
free to rotate around each other. Comparison of this model with 
force--extension curves for single-stranded DNA shows that a better fit
is obtained from a Modified FJC, 
\beq\label{FJC}
l_{MFJC}(f) =  d \left( 1 + \frac{f}{\g_{ss}} \right) \times l_{FJC} (f)
\eeq
which takes into account elasticity effects on the rod length.
Standard fit parameters are
$d = 0.56$ nm, $b=1.4$ nm, and $\g_{ss}=800$ pN.
\item {\it Extensible Worm-Like Chain (WLC) model:}
\beq\label{DS}
l_{WLC}(f) = L \left[ 1 -\frac12 \left(\frac{k_BT}{f A}\right)^{1/2} +
  \frac{f}{\g_{ds}} \right] 
\eeq
is the formula for the high-force extension of an elastic chain with 
persistence length equal to $A$. Experiments shows it is an excellent 
description of double-stranded DNA at high forces, with $L = 0.34$ nm, $A = 48$ nm and 
$\g_{ds} = 1000$ pN. 
\end{itemize}

\subsubsection{Free-energy landscape for the sequence}
\label{sec:GnB}

\begin{table}
\begin{center}
\begin{tabular}{|c|c|c|c|c|} \hline $g_0$ &A&T&C&G \\ \hline A &1.78
&1.55 &2.52 & 2.22\\ \hline T &1.06 &1.78 &2.28 &2.54 \\ \hline C&
2.54 &2.22 &3.14& 3.85 \\ \hline G & 2.28& 2.52 & 3.90&3.14\\ \hline
\end{tabular}
\end{center}
\caption{Binding free energies $g_0(b_i,b_{i+1})$ (units of $k_B T$)
obtained from the MFOLD server \cite{Zuk,San} for DNA at room
temperature, pH=7.5, and ionic concentration of 0.15 M. The base
values $b_i, b_{i+1}$ are given by the line and column respectively.}
\label{tableg0}
\end{table} 
  
Let $b_i=A,T,C$, or $G$ denote the
$i^{\rm th}$ base along the $5'\to3'$ strand (the other strand is
complementary), and ${B}=\{b_1,b_2,\ldots , b_N\}$.  The free energy
excess when the first $n$ bp of the molecule are open with respect to
the closed configuration ($n=0$) is~\cite{Coc4}
\begin{equation} \label{p} 
G(n;{B})= \sum_{i=1} ^n g_0(b_i,b_{i+1})\ .
\end{equation} 
$g_0(b_i,b_{i+1})$ is the binding energy of base-pair (bp)
number $i$; it depends on $b_i$ (pairing interactions) and on the
neighboring bp $b_{i+1}$ due to stacking interactions.  $g_0$ is
obtained from the MFOLD server \cite{Zuk,San}, and listed in
Table~\ref{tableg0} for a 150 mM NaCl, room temperature and pH 7.5. The values of the free-energies should be changed for different ionic conditions and temperatures.

\begin{figure}
\includegraphics[height=6cm]{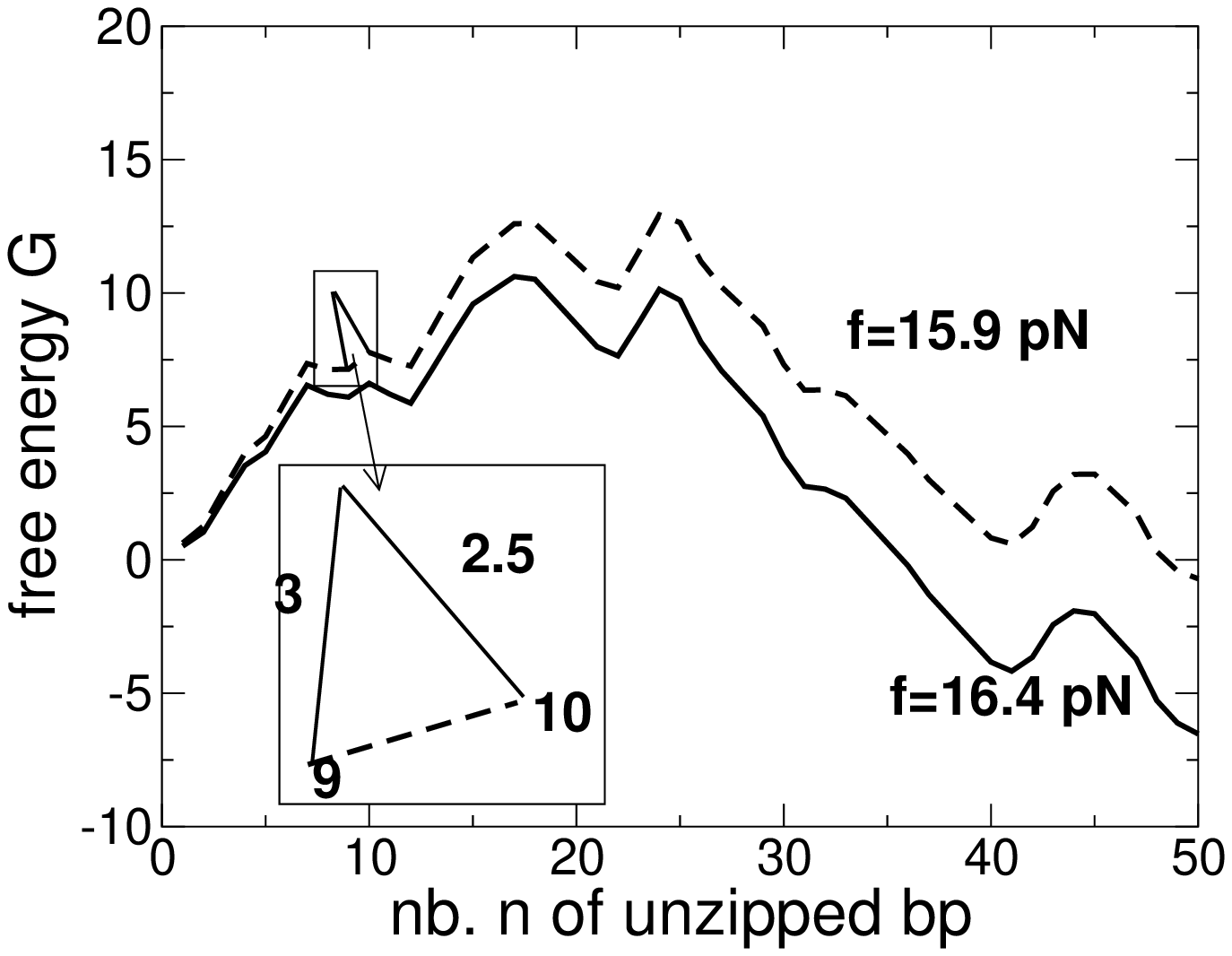}
\caption[]{Free energy $G$ (units of k$_B$T) to open the first $n$
base-pairs, for the first 50 bases of the DNA $\lambda$--phage at
forces 15.9 (dashed curve) and 16.4 pN (full curve). For $f=15.9$ pN
the two minima at bp 1 and bp 50 are separated by a barrier of 12
k$_B$T. Inset: additional barrier representing the dynamical rates
(\ref{ratemd}) to go from base 10 to 9 (barrier equal to $2 g_{ss}$=2.5
k$_B$T), and from base 9 to 10 (barrier equal to $g_0(b_9,b_{10})$=3
k$_B$T), see text. }
\label{gf50}
\end{figure}

As an illustration we plot the free energy $G(n;\Lambda )$ of the first 50
bases of the $\lambda$ phage sequence, $\Lambda = (\lambda _1, \lambda
_2, \ldots ,\lambda_N)$, in Figure~\ref{gf50} after subtraction of
$n \;g_{ss}(f)$  for forces $f=15.9$ and $16.4$ pN. 
$g_{ss}(f)$ is the work to stretch the
two opened single strands when one more bp is opened, and calculated from
the Modified FJC model (\ref{FJC}).
The substraction allows to compare the increase in the free-energy due to 
the opening of the sequence to the gain resulting from the release of
ssDNA polymers at a given force.

At these forces the two global minima in Figure~\ref{gf50} are
located in $n=1$ (closed state) and $n=50$ (partially open state).
Experiments on a small RNA molecule, called P5ab \cite{Lip01}, have been
performed at the critical force $f_c$ such that the closed state has
the same free energy than the open one: $G(N;\Lambda )=N\; g_{ss}(f_c)$. 
They showed that, as the
barrier between these two minima is not too high, the molecule
switches between these two states, see Section \ref{direct}.

\subsection{Fluctuations at equilibrium}

\subsubsection{Case of a single polymer}
\label{sec:manip}
We now consider the orders of magnitude of the fluctuations of the polymer. 
When submitted to a force of $f = 15$ pN,
the average extension of the polymer is $\bar x = n x^m$ with
$x^m = l(f)$. We use for single-stranded DNA the MFJC model,
and for double-stranded DNA the WLC model, with the parameters
discussed in section~\ref{poly-models}; then
we get
$x^m_{ss} = 0.46$ nm and $x^m_{ds}=0.33$ nm for the ss- and ds-DNA.
At thermal equilibrium the extension will fluctuate around these average
values. The fluctuations are controlled by the {\it microscopic effective 
spring constant} $k^m(l) = w''(l)=1/l'(f)$.
For ds- and ss-DNA we find, respectively, $k^m_{ds} = 1311$ pN/nm 
and $k^m_{ss} = $138 pN/nm according to the models above.
For a polymer with $n$ monomers the stiffness is $k = k^m/n$ since the 
effective spring constant is given by 
$k(x,n) = \frac{\partial^2}{\partial x^2} W(x,n) = k^m(x/n)/n$.

Alternatively the force $f$ exterted on the polymer will fluctuate 
around its average value $\bar f$ if 
its extremities are kept at a fixed distance $x$ from each other.
These fluctuations of force  (in the fixed-extension ensemble) 
and extension (in the fixed-force ensemble) are easily computed by a quadratic
expansion of the free-energy around the average
\ie when approximating the polymer with a spring of stiffness $k^m/n$, with the
result
\beq\label{fluctsolo}\begin{split}
&\la \d x^2 \ra = \frac{k_BT}{k^m} n \ , \hskip1cm
\la \d f^2 \ra = \frac{k_BT\; k^m}{n} \ ,
\end{split}\eeq
Defining  $\d \bar x = \sqrt{\la \d x^2 \ra}$ and
$\d \bar f = \sqrt{\la \d f^2 \ra}$
we get
\beq\label{fluc1}\begin{split}
&\frac{\d \bar x}{\bar x} = \sqrt{\frac{k_BT}{k^m (x^m)^2}}\frac{1}{\sqrt{n}} \ , \hskip1cm
\frac{\d \bar f}{\bar f} = \sqrt{\frac{k_BT\; k^m}{\bar f^2}}
\frac{1}{\sqrt{n}} \ .
\end{split}\eeq
As expected the relative fluctuations of both force
 and extension become smaller
and smaller as the number $n$ of monomers increases. 
Some values are reported in Tables \ref{tab:II} and \ref{tab:III}.

\begin{table}[t]
\begin{tabular}{|c|cc|c|}
\hline
$n$ & $\d \bar x/\bar x$ & $\d \bar f/\bar f$  & $\t$ (s) \\
\hline
10 & 0.117 & 0.496 & $4.8 \cdot 10^{-9}$ \\
40 & 0.058 & 0.248 & $7.7 \cdot 10^{-8}$ \\
100 & 0.037 & 0.157 & $4.8 \cdot 10^{-7}$ \\
400 & 0.018 & 0.078 & $7.7 \cdot 10^{-6}$ \\
1000 & 0.012 & 0.050 & $4.8 \cdot 10^{-5}$ \\
\hline
\end{tabular}
\caption{
Fluctuations of single stranded DNA at $f=15$ pN and $T=16.7 \, ^\circ$C;
$\d \bar x/\bar x = 0.37/\sqrt{n}$,
$\d \bar f/\bar f = 1.57/\sqrt{n}$,
$\tau = 4.83 \cdot 10^{-11} \text{ s } n^2$.
}
\label{tab:II}
\end{table}

\begin{table}[t]
\begin{tabular}{|c|cc|c|}
\hline
$n$ & $\d \bar x/\bar x$ & $\d \bar f/\bar f$   & $\t$ (s)  \\
\hline
100 & 0.017 & 0.483 &  $5.1 \cdot 10^{-8}$    \\
400 & 0.0085 & 0.241 &  $8.1 \cdot 10^{-7}$    \\
1000 & 0.0054 & 0.153 & $5.1 \cdot 10^{-6}$ \\
4000 & 0.0027 & 0.076 & $8.1 \cdot 10^{-5}$ \\
10000 & 0.0017 & 0.048 & $5.1 \cdot 10^{-4}$ \\
\hline
\end{tabular}
\caption{
Fluctuations of double stranded DNA at $f=15$ pN and $T=16.7 \, ^\circ$C;
$\d \bar x/\bar x = 0.17/\sqrt{n}$,
$\d \bar f/\bar f = 4.83/\sqrt{n}$,
$\tau = 5.1 \cdot 10^{-12} \text{ s } n^2$.
}
\label{tab:III}
\end{table}

\subsubsection{Case of several polymers (fixed-distance setup)}

Consider now the case of several polymers e.g. linker and open part of 
the molecule attached one after the other. In a fixed-force experiment
the components of the setup are independent (at the level
of the saddle-point approximation), and the fluctuations in the extensions 
simply add up. In the fixed-distance setup, however, correlations
between the extension make the analysis more complicated.

As a concrete example we consider the setup in figure~\ref{fig:setup}A.
The linker joining $x_1$ and $x_2$ 
is a double-stranded DNA segment of $N_{ds}$
bases. The two linkers joining $(x_2,x_3)$ and $(x_3,x_4)$ are
single-stranded DNA segments of $N_{ss} = 
N^0_{ss} + n$ bases, where $n$ is the
number of opened base-pairs.
The centers of the two optical traps are in $0$ and $X$.
We call $x_1$ the position of the first bead, and $x_4$ the
position of the second.
The probability $P_{eq}(n,x_1,x_2,x_3,x_4) = e^{-F/k_BT}$ 
where the free energy $F$ reads
\beq\label{Fmanip}
\begin{split}
F(\vec x,n)&= \frac12 k_1 x_1^2 + W_{ds}(x_2-x_1,N_{ds}) \\
&+W_{ss}(x_3-x_2,N_{ss}) 
 + W_{ss}(x_4-x_3,N_{ss})\\& + \frac12 k_2 (x_4 - X)^2 + G(n;B) \ . \\
\end{split}\eeq
where
$W_{ds}(x,N_{ds})=N_{ds} w_{ds}(x/N_{ds})$ 
and $W_{ss}(x,N_{ss})=N_{ss} w_{ss}(x/N_{ss})$ 
are the elongation free energies of the double strand and single strand 
respectively.

In order to study the fluctuations in this setup
we first find the maximum of $P_{eq}$ assuming that $G(n;B) = n \;g_0$
\ie a uniform sequence $B$, and treating $n$ as a continuous variable
assuming that it is large.
At the maximum $x_i = \bar x_i$ and we define
\beq\label{barl}\begin{split}
&x^m_{ds} = \frac{\bar x_2-\bar x_1}{N_{ds}} \ , \\
&x^m_{ss} = \frac{\bar x_3-\bar x_2}{N_{ss}} =
\frac{\bar x_4-\bar x_3}{N_{ss}} \ . \\
\end{split}\eeq
The saddle-point condition $\partial_{x_i} F_A = 0$ 
gives the following equations, that represent the force balance
condition along the chain:
\beq\label{esaddle}
\begin{split}
k_1 \bar x_1 = w'_{ds}(x^m_{ds}) = w'_{ss}(x^m_{ss}) 
= k_2 (X-\bar x_4) \equiv \bar f \ .
\end{split}
\eeq
The derivative with respect to $n$ gives,
using Eqs.~(\ref{esaddle}) and (\ref{formulas-poly}), the condition
\beq\label{fsaddle}
g_0 = 2 [ x^m_{ss}w'_{ss}(x^m_{ss})-w_{ss}(x^m_{ss}) ]
= g_{ss}(\bar f) \ ,
\eeq
that allows to find the force $\bar f$ transmitted along the chain.
Once (\ref{fsaddle}) is solved, 
the extension of the beads and of the double and single stranded
part of DNA ($\bar x_1$, $X-\bar x_4$, $x^m_{ds}$ and $x^m_{ss}$ 
respectively) are determined by Eq.~(\ref{esaddle}).
Finally the number of open bases $\bar n$ is determined
by
\beq\label{barn}
\bar x_1 + N_{ds} x^m_{ds} + 2 (N^0_{ss} + \bar n) x^m_{ss} + 
(X - \bar x_4) = X \ .
\eeq
Note that the value of $\bar f$ is determined only by $g_0$.

We work at temperature $T=16.7 \, ^\circ$C ($k_B T=4$ pN nm) 
and choose a uniform molecule with $g_0 = 2.69 \, k_B T$, which
is a representative value for the pairing free energies in
table~\ref{tableg0}. We use
the same models as in section~\ref{sec:manip} 
for the single and double stranded DNA, with $N_{ds}=3120$ and
$N^0_{ss}=40$.
Then solving Eq.(\ref{fsaddle}) we get $\bar f = 16.5$ pN, 
and from Eq.(\ref{esaddle}) we get
$x^m_{ss}=0.47$ nm, $x^m_{ds}=0.33$ nm.
We choose $k_1 = 0.1$ pN/nm, then 
$\bar x_1 =165$ nm, and $k_2 = 0.512$ pN/nm, then 
$X - \bar x_4 =32$ nm.
Given these values,
$\bar n$ is defined by $X$ using Eq.~(\ref{barn}):
\beq
\bar n = \frac{X-1264}{0.94} \ ,
\eeq
with $X$ expressed in nanometers.

\begin{table}[t]
\begin{tabular}{c|c|c|c}
\hline
$X$ & $\bar n$ & $k_{\rm eff} $  
& $\sqrt{\langle \d n^2 \rangle}$  \\
\hline
1273 & $10^1$ &  0.067 & 8.2 \\
1358 & $10^2$ &  0.062 &  8.5 \\
2204 & $10^3$ &  0.036 &  11.2 \\
10664 & $10^4$ & 0.0068 & 25.7 \\
\hline
\end{tabular}
\caption{Saddle-point calculation for the setup in figure~\ref{fig:setup}A with
a uniform molecule and 
$k_1 = 0.1$ pN/nm, $k_2 = 0.512$ pN/nm, $N_{ds}=3120$, $N^0_{ss}=40$.
The force along the molecule is $\bar f = 16.5$, then $k^m_{ss}=152$ pN$/$nm,
$k^m_{ds}= 1416$ pN$/$nm and $k^s_{\rm eff} =0.07$ pN$/$nm.
}
\label{tab:I}
\end{table}

For the same setup we
can compute the fluctuations of $n$ and of the elongations
of the elements of the setup.
In particular the fluctuations of the bead positions are measurable 
in the experiment.

Let us define $\d x_i = x_i - \bar x_i$ and
$\d n = n - \bar n$.
To simplify the formalism we also
define
$\d x_{ds} = \d x_2- \d x_1$,
$\d x_{ss}^L = \d x_3 - \d x_2$, and
$\d x_{ss}^R = \d x_4 - \d x_3$.
A quadratic expansion of $F$ around its minimum gives
\beq\begin{split}
\delta & F \sim 
 \frac12 k_1 \d x_1^2 + \frac12 k_2 \d x_4^2 +
\frac{w''_{ds}(x^m_{ds})}{2N_{ds}} \d x_{ds}^2 \\
&+\frac{ w''_{ss}(x^m_{ss})}{2N^0_{ss}+\bar n}
[(\d x_{ss}^L- x^m_{ss} \d n)^2 + 
(\d x_{ss}^R - x^m_{ss} \d n )^2 ] \ .
\end{split}\eeq
Using (\ref{FJC}) and (\ref{DS})
we get $k^m_{ss}=w''_{ss}(x^m_{ss})=152$ pN$/$nm and 
$k^m_{ds}=w''_{ds}(x^m_{ds}) = 1416$ pN$/$nm.

One should take care of the fact that
$\d x_1 + \d x_4 + \d x_{ds} + 
\d x_{ss}^L + \d x_{ss}^R = 0$; it is convenient
to express $\d x^R_{ss}$ as a function of the others
since its fluctuations are identical to the ones of
$\d x^L_{ss}$.
The quadratic expansion of the function $\delta F$ has the form
$\delta F = \frac12 \d\xx A \d\xx$
where $\d\xx = (\d x_1,\d x_4,\d x_{ds},\d x^L_{ss},x^m_{ss}\d n)$ 
and 
\beq\begin{split}
A &= \frac{k^m_{ss}}{N^0_{ss}+\bar n} \; 
\begin{pmatrix}
1 & 1 & 1 & 1 & 1 \\
1 & 1 & 1 & 1 & 1 \\
1 & 1 & 1 & 1 & 1 \\
1 & 1 & 1 & 2 & 0 \\
1 & 1 & 1 & 0 & 2 \\
\end{pmatrix} \\
&+ 
\begin{pmatrix}
k_1 & 0 & 0 & 0 &  0 \\
0 & k_2 & 0 & 0 & 0\\
0& 0 & k^m_{ds}/N_{ds} & 0 &0 \\
0 & 0 & 0 & 0 & 0\\
0 &0 & 0 & 0 &0 \\
\end{pmatrix}
\end{split}\eeq
The inverse of the matrix $A$ is
\beq
A^{-1} = 
\begin{pmatrix}
\frac{1}{k_1} & 0 & 0 & -\frac{1}{2 k_1} &-\frac{1}{2 k_1}  \\
0 & \frac{1}{k_2} & 0 & -\frac{1}{2 k_2} & -\frac{1}{2 k_2} \\
0 & 0 & \frac{N_{ds}}{k_{ds}^m} & -\frac{N_{ds}}{2k_{ds}^m} &-\frac{N_{ds}}{2k_{ds}^m} \\
-\frac{1}{2 k_1} & -\frac{1}{2 k_2} & -\frac{N_{ds}}{2k_{ds}^m} & \frac{1}{4k_{\rm eff}} & \frac1{4 k_{\rm eff}^s} \\
-\frac{1}{2 k_1} & -\frac{1}{2 k_2} & -\frac{N_{ds}}{2k_{ds}^m} & \frac1{4k_{\rm eff}^s} & \frac{1}{4 k_{\rm eff}} \\
\end{pmatrix}
\label{tab}
\eeq
where
\beq\begin{split}
&\frac1{ k^s_{\rm eff}}=
\frac{1}{k_1} + \frac1{k_2} +\frac{N_{ds}}{k^m_{ds}} \ , \\
&\frac1{ k_{\rm eff}}=\frac1{k^s_{\rm eff}}+
2 \frac{N^0_{ss}+\bar n}{k^m_{ss}} \ .
\end{split}\eeq
This immediately gives
\beq\begin{split}
k_B T (A^{-1})_{1,1}=&\langle \d x_1^2 \rangle = \frac{k_B T}{k_1} \\
k_B T (A^{-1})_{2,2}=&\langle \d x_4^2 \rangle = \frac{k_B T}{k_2} \\
k_B T (A^{-1})_{3,3}=&\langle \d x_{ds}^2 \rangle = \frac{k_B T N_{ds}}{k^m_{ds}} \\
k_B T (A^{-1})_{4,4}=&\langle (\d x^L_{ss})^2 \rangle = \frac{k_B T}{4 k_{\rm eff}} \\
\frac{k_B T}{(x^m_{ss})^2} (A^{-1})_{5,5}=&\langle \d n^2 \rangle =\frac{k_B T}{4 k_{\rm eff}(x^m_{ss})^2}
\end{split}
\label{tabinv}
\eeq
and shows that the fluctuations of $n$ are dominated by
the weakest element of the setup; moreover the correlation between the 
beads displacements $\d x_1$, $\d x_4$ and the fluctuations of
the number of open base-pairs $\d n$ is
 $\langle \d n \d x_{1} \rangle= - \frac{k_B T}{2 k_{1} x^m_{ss}}$
and $\langle \d n \d x_{4} \rangle= - \frac{k_B T}{2 k_{2} x^m_{ss}}$;
the stiffer the optical trap , the weaker is the correlation between the 
location of the  bead and the number of open bases.
Examples are given in Table \ref{tab:I}.

\subsection{Effective dynamical models}
\label{direct}

In the simplest dynamical models the fork (separating the open and closed 
portions of the molecule) undergoes a biased random motion in the
sequence landscape. The linkers are treated at equilibrium, which is
correct if their characteristic time scales are much smaller than the
average time to open or close a base-pair.

\subsubsection{Time scales for the polymeric components of the setup}
\label{ts}

In this section we recall the typical time scales of the polymeric
components in the setup. Assume that the polymers are subject to
a Brownian force $\eta(t)$ which is a zero-average Gaussian 
process with autocorrelation function
$\langle\eta(t)\eta(0)\rangle=2\Gamma T \delta(t)$.
Let $\Gamma$ be the friction coefficient of the polymer 
\cite{DoiEdwards}, that is, the ratio between the viscous force exerted by
the solvent and the velocity. As will be shown in Section \ref{sec:model} 
the friction coefficient scales as $\G = \g^m n/3$ with 
$\g^m_{ss} = \g^m_{ds} \sim  2 \cdot 10^{-8}$ pN s/nm$^2$.
Then, approximating $f(x,n) \sim k^m x/n$,
the relaxation time for an isolated polymer of $n$ bases
is given by
\beq\label{tau3}
\t = \frac{\g^m n^2}{3 k^m} \ .
\eeq
Note that the factor 3 at the denominator of the above equation is an
approximation for the true factor $\pi^2/4$. The validity of its approximation,
and the simplification it leads to will be discussed in 
Appendix \ref{app:poly}.

It is useful to compare the amplitude of the force fluctuations with 
the noise. To do this we approximate $\la \d f(t) \d f(0) \ra \sim 2 
\t \la \d f^2 \ra \d(t) =
2 T \G_f \, \d(t)$. Then, using Eq.~(\ref{fluctsolo}) to estimate 
$\la \d f^2 \ra$, 
we get $\G_f = n \g^m/3 = \G$, and (not surprisingly)
the force fluctuations are of the same order of the noise term.

From Table \ref{tab:II} the relaxation time of the unzipped strands is smaller 
that the typical base-pair opening (or closing) time as long as the number
$n$ of unzipped bases is smaller than a few hundreds. This is the case
in particular for unzipping experiments on short RNA molecules.

\subsubsection{Random walk in the sequence landscape} 

Let us first model the motion of the fork alone, that is, assuming that
the other components of the setup are at equilibrium. 
We consider a DNA molecule 
unzipped under a fixed force $f$ in the sequence-landscape 
$G(n; B)- n \; g_{ss}(f)$ of Figure \ref{gf50}. 
The fork, whose position is
denoted by $n(t)$, can move forward ($n\to n+1$) or backward 
($n\to n-1$) with rates (probability per unit of time) equal to, respectively,
\begin{equation}\begin{split}
&r_o(b_{n+1},b_{n+2})= r\; \exp \big[
-\b g_0(b_{n+1},b_{n+2})\big] \ , \\
&r_c= r\; \exp \big[{-2 \b g_{ss} (f)}\big] \ ,
\label{ratemd}
\end{split}\end{equation} 
where $\b=1/k_B T$, see Fig \ref{gf50}. 
The value of the attempt frequency
$r$ is of the order of $10^{6}$ Hz \cite{Coc4,Man07,Boc04}. 
Expression (\ref{ratemd}) for the rates is derived from the
following assumptions. First the rates should satisfy detailed
balance.  Secondly we impose that the opening rate $r_o$ depends on
the binding free energy, and not on the force, and vice-versa for the
closing rate $r_c$. This choice is motivated by the fact that the
range for base-pair interaction is very small: the hydrogen and
stacking bonds are broken when the bases are kept apart at a fraction
of Angstrom, while the force work is appreciable on the distance of the
opened bases ($\approx 1$ nm).  On the contrary, to close the base
pairs, one has first to work against the applied force, therefore the
closing rate $r_c$ depends on the force but not on the sequence. This
physical origin of the rates is reported in the inset of
Figure~\ref{gf50}. Notice that, as room temperature is much smaller than
the thermal denaturation temperature, we safely discard the existence
of denatured bubble in the zipped DNA portion.

\begin{figure}
\includegraphics[height=6cm,angle=-90]{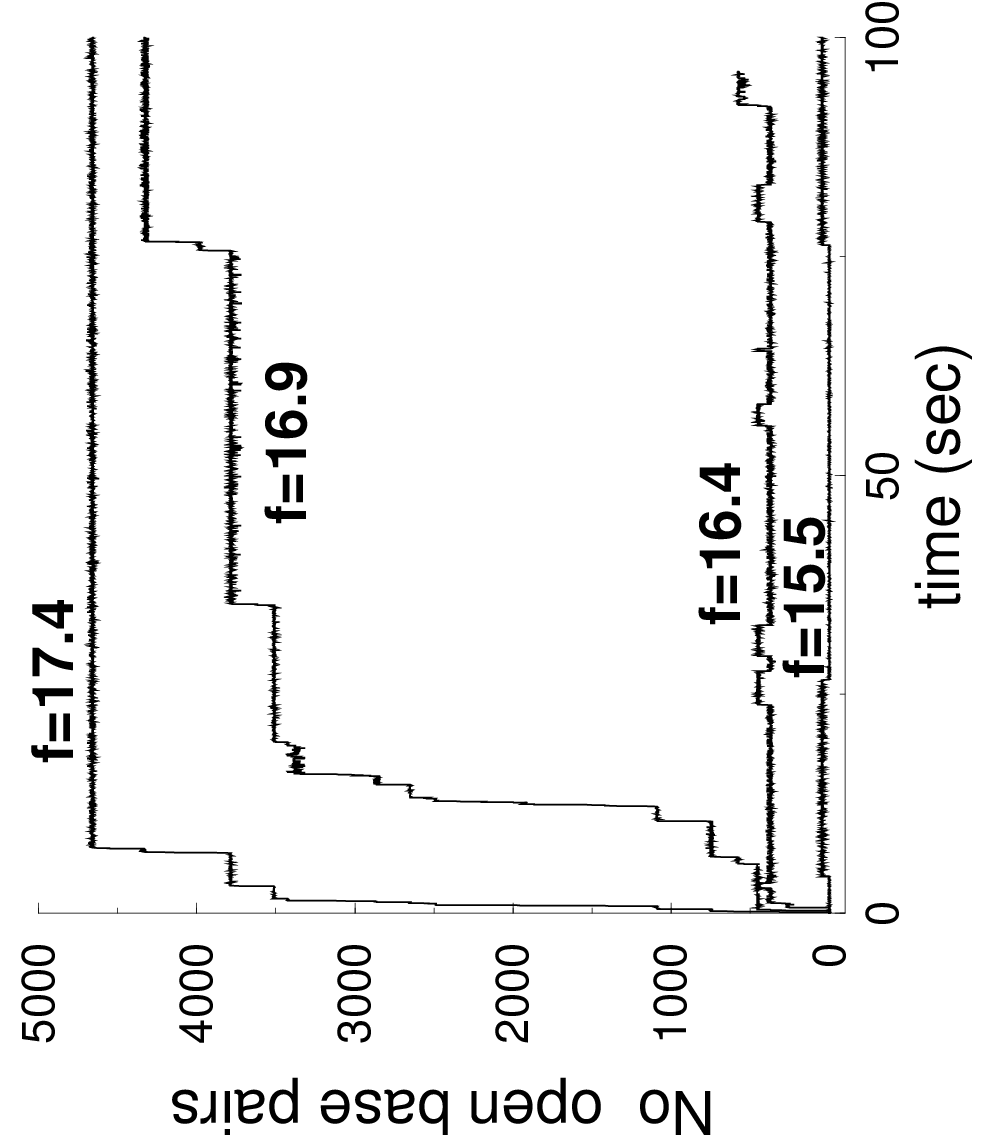}
\caption{Number of open base-pairs as a function of the time for
various forces (shown on Figure).  Data show one numerical unzipping
(for each force) obtained from a Monte Carlo simulation of the random
walk motion of the fork with rates (\ref{ratemd}).}
\label{dinf15.5}
\end{figure}

An example of unzipping dynamics for the $\lambda$-phage sequence is shown
in Figure~\ref{dinf15.5}. The characteristic pauses in the unzipping, 
present in experiments and
corresponding to deep local minima in the sequence-landscape are
reproduced.  The rates (\ref{ratemd})
lead to a master equation for the probability $\rho_n(t)$ for the
fork to be at site $n$ at time $t$
\begin{equation} \frac{d\rho_n(t)}{dt}= -\sum_{m=0}^{N}
T_{n,m}\;\rho_m(t)
\end{equation} where the matrix $T_{n,m}$ is tridiagonal with nonzero
entries $T_{m-1,m}=-r_c(f)$, $T_{m+1,m}=-r_o(m)$ and
$T_{m,m}=r_o(m)+r_c(f)$. Given this transition matrix the opening
dynamics can be simulated with a Monte Carlo.  For small RNA or DNA
molecules the transition matrix $T_{n,m}$ can be diagonalized
numerically \cite{Coc4}.  The smallest non--zero eigenvalue gives the
switching time between a closed and open configuration for hairpin
with a free energy barrier such as the one plotted in
Figure~\ref{gf50}.

\subsubsection{Dynamics of the bead with equilibrated linkers and strands}
\label{IIC3}

In a typical experiment the force is exerted on the molecule through the action of
a (magnetic or optical) trap on the bead. While the external 
force on the bead
can be considered as constant (e.g. in a magnetic trap), the force acting 
on the fork fluctuates unless the trap (and the molecular
construction) is very soft, see Eq.~(\ref{fluc1}). Therefore the fixed-force
model of the previous section has to be modified. 
In addition
the bead, of size $R \simeq 1\ \mu$m, is a slow component whose
dynamics need to be taken into account. Let us denote by $k$ the stiffness of
the trap and $\gamma$ the friction of the bead in the solvent of 
viscosity $\eta$. Typical values for these quantities are $k=0.1-0.5$ pN/nm, 
and $\gamma = 6 \pi R \eta = 1.67 \; 10^{-5}$~pN~s/nm. 
The characteristic relaxation time of the
bead is thus $\tau = \gamma/k\simeq 0.2-1$~ms.

The coupled dynamics of the fork and the bead was considered by
Manosas et al. \cite{Man07} in the case of small RNA unzipping, with
a single optical trap.
For such small molecules the relaxation time of the unzipped strands is 
expected to be much 
smaller than the characteristic time of the bead, and
the molecule can be considered at equilibrium.
The dynamical scheme therefore consists in a coupled evolution equation for 
the location of the bead and of the fork. The bead position obeys a 
Langevin equation including the external force and the force exerted by the
fork through the (equilibrated) linkers and unzipped strands, while
the fork moves with rates (\ref{ratemd}) with a bead location-dependent force.

A main conclusion of Ref.~\cite{Man07} is that, in the absence of feedback
imposing a fixed force on the molecule,
 the trap stiffness must be as low as
possible to detect jumps between closed and open configurations of the
RNA molecule. We will discuss the validity of this statement in an 
information--theoretic setting in section \ref{info}.

\section{Dynamical modelling of the setup and its components}
\label{sec:model}

The assumption that the linkers and the unzipped strands are at 
equilibrium as the unzipping proceeds is correct for short molecules
as was the case in \cite{Man07}. For long DNA molecules the relaxation 
time of the unzipped strands may become large, and a dynamical modelling of the
polymers involved in the molecular construction cannot be avoided.

The purpose of this section is to describe how such a dynamical
model can be implemented.
We hereafter denote by ``setup'' the full molecular construction
that is used in a given experiment, including linkers, beads, etc. while
the word ``molecule'' refers to the part of DNA which has
to be opened. 
In an idealized description, the state variable is a vector 
$\vec{x} = (x_1, \cdots, x_p)$ whose elements are the
distances from a reference position (that can be either the center
of an optical trap, or a fixed `wall' to which the polymers are
attached) of the extremities of the polymeric components in the setup.
In addition to $\vec x$,
the number of open base-pairs $n$ is needed to complete the description
of the state of the setup.

As discussed in section \ref{thermo} the total free energy $F(\vec x,n)$ 
of a setup is the sum of different contributions coming from all
the elements of the setup. A typical example is given in Eq.~(\ref{Fmanip}).

Our aim is thus to construct a dynamical model that holds
on intermediate time scales, $t \gtrsim 10^{-6}$ s, and
\begin{enumerate}
\item gives the correct equilibrium Gibbs measure 
$P_{eq}(\vec x,n) = \exp(-F(\vec x,n)/(k_B T))$,
\item reproduces the relaxation times for the different 
elements of the setup, as discussed below, 
\item gives reasonable dynamical correlations between
different elements of the setup.
\end{enumerate}
It is worth to stress at this point that ours is a coarse-grained 
model which does not take into account the motion of the individual 
monomers. It is expected that the dynamics on time-scales smaller than
the typical sojourn time of the fork on a base $(\gtrsim 10^{-6}$ s) 
is not relevant to our study of unzipping.

\subsection{Langevin dynamics for the polymers and the beads}

First we consider the dynamics of $\vec x$ at fixed $n$.
In Appendix~\ref{app:poly} we show that for long enough times
the dynamics of the setup can be described by a system of coupled
Langevin equations:
\beq\label{text:finalchain}
\G_{ij} \dot{x_j} = -\frac{\partial F}{\partial x_i} + \eta_i \ ,
\eeq
where $i,j = 1,\cdots,p$ and:
\begin{itemize}
\item the free energy $F(\vec x)$ is the sum of a contribution
coming from each element of the setup:
\begin{itemize}
\item each optical trap contributes $\frac12 k \D x^2$, where
$\D x$ is its elongation; 
\item a bead in position $i$ subjected to a constant force 
gives a contribution $-f x_i$;
\item a polymer gives a contribution $W_i(\D x,N_i)$, with $\D x$
its elongation and $N_i$ its number of monomers.
\end{itemize}
For example, the total 
free energy of the setups in figure~\ref{fig:setup} are
\beq\begin{split}
F_A(\vec x)&= \frac12 k_1 x_1^2 + W_{ds}(x_2-x_1,N_{ds})
+W_{ss}(x_3-x_2,N_{ss}) \\
& + W_{ss}(x_4-x_3,N_{ss}) + \frac12 k_2 (x_4 - X)^2 \ , \\
F_B(\vec x)&= W_{ds}(x_1,N_{ds})
+W_{ss}(x_2-x_1,N_{ss}) \\ &+ W_{ss}(x_3-x_2,N_{ss}) - fx_3 \ . 
\end{split}\label{fafb}
\eeq
\item $\vec \eta$ is a Gaussian white noise with zero average and
variance $\la \h_i(t) \h_j(0) \ra = 2 k_BT \G_{ij} \d(t)$, as requested
by the fluctuation-dissipation relation;
\item the matrix $\G$ is a tridiagonal matrix such that:
\begin{itemize}
\item the diagonal element $\G_{ii}$ is the sum of three contributions:
\begin{itemize} 
\item a term
$\g^m_{i-1} N_{i-1}/3+\g^m_i N_i/3$
coming from the adjacent polymers (if any);
\item a term $\g$ coming from the bead (if any) attached to $x_i$;
\item
a term taking into account the viscosity of the
$N_c$ base-pairs of the DNA molecule attached to the fork
($x_3$ and $x_2$ in figure~\ref{fig:setup}A and B 
respectively) that are not open. This term has the Fleury form 
$\g_{mol} = \g' N_c^{3/5}$ and has to be
added to the diagonal element of $\G$ corresponding to the
fork position;
\end{itemize}
\item the offdiagonal elements are zero, except 
$\G_{i,i+1} = \G_{i+1,i} = \g_{i+1}^m \frac{N_{i+1}}6$ that get
a contribution from the polymer joining $x_i$ and $x_{i+1}$.
\end{itemize}
For instance the setups in figure~\ref{fig:setup} correspond
to the matrices:
\beq\begin{split}
&\G_B = 
\begin{pmatrix}
 \g_{ds}^m \frac{N_{ds}}3+\g_{ss}^m \frac{N_{ss}}3 &
\g_{ss}^m \frac{N_{ss}}6 & 0 \\
\g_{ss}^m \frac{N_{ss}}6 & 2 \g_{ss}^m \frac{N_{ss}}3 + \g' N_c^{3/5} & 
\g_{ss}^m \frac{N_{ss}}6 \\
0 & \g_{ss}^m \frac{N_{ss}}6 & \g + \g_{ss}^m \frac{N_{ss}}3 \\
\end{pmatrix} \ , \\
&\G_A = 
\begin{pmatrix}
\g + \g_{ds}^m \frac{N_{ds}}3 & 
\begin{matrix} \g_{ds}^m \frac{N_{ds}}6 & 0 & 0 \end{matrix} \\
\begin{matrix}
\g_{ds}^m \frac{N_{ds}}6 \\
0 \\
0 \\
\end{matrix}
& \G_B
\end{pmatrix} \ . \\
\end{split}\label{gammaagammab}
\eeq
\end{itemize}

A detailed derivation of these results and in particular
of the form of the matrix $\G$ can be found in 
Appendix~\ref{app:poly}.

\subsection{Fork dynamics}
\label{sec:fork}

The Langevin equation for the polymer dynamics at fixed $n$ 
must be complemented
with transition rates for the dynamics of $n$.
To this aim we discretize the Langevin equation with time
step $\D t$, and
at each time step we allow the opening $n \to n+1$
or closing $n \to n-1$ of a base-pair at most.

The dynamics takes the form of a discrete time Markov
chain, with transitions 
$(\vec x,n) \to (\vec x', n')$ and $n' \in \{n, n\pm 1\}$.
The total free energy $F(\vec x,n) = F_{\rm setup}(\vec x,n) +
G(n;B)$, where the first contribution has been discussed in
the previous section and $G(n;B)$ is the pairing free energy
of the molecule, as discussed in section~\ref{sec:GnB}.
In Appendix~\ref{app:bd} we show that in order to satisfy
the detailed balance condition with respect to 
$P_{eq}(\vec x,n) = \exp(-F(\vec x,n)/(k_B T))$, one should
perform a single step following the procedure:
\begin{enumerate}
\item Choose 
whether to stay ($n'=n$), 
to open ($n'=n+1$) or to close ($n'=n-1$) a base, 
with rates $r^{s,o,c}(\vec x,n)$ respectively:
\beq\label{ratesZ}
\begin{split}
r^o(\vec x,n) &= r \D t e^{\b [G(n;B)-G(n+1;B)]} \ , \\
r^c(\vec x,n) &= r \D t e^{\b F(\vec x,n)-\b F(\vec x,n-1)} \ , \\
r^s(\vec x,n) &= 1 -r^o(\vec x,n)-r^c(\vec x,n) \ .
\end{split}\eeq
\item If the choice was to open, {\it first} perform a discrete
Langevin step $\vec x \to \vec x'$
at fixed $n$ and {\it then} increase $n$ by one.
\item If the choice was to close, 
{\it first} decrease $n$ by one and 
{\it then} perform a discrete Langevin step $\vec x \to \vec x'$
at fixed $n'=n-1$.
\item If the choice was to stay, just perform a discrete Langevin step $\vec x \to \vec x'$ 
at fixed $n$.
\end{enumerate}

The Langevin equation is discretized in a standard way by integrating
Eq.~(\ref{text:finalchain}) over a time $\D t$:
\beq
x_i(t+\D t) = x_i(\D t) + \G^{-1}_{ij} 
\left[ -\frac{\partial F(\vec x)}{\partial x_j} 
\D t + E_j \right] \ ,  
\eeq
where $E_j = \int_0^{\D t} \eta_j(t) dt$ are 
Gaussian variables with zero average and variance
\beq
\la E_i E_j \ra = 2 k_BT \G_{ij} \D t
\eeq
that are independently drawn at each discrete time step.

\subsection{Free energy at finite $n$}

In section~\ref{thermo} we discussed some models for the free 
energy $W(x,n)$ of a polymer with $n$ monomers and extension $x$.
In the limit $x,n \to\io$ at fixed extension per monomer, $l=x/n$, the
free-energy enjoys an extensivity property: $W(x,n)= n\; w(l)$.
However, in our simulations we might be interested to regimes where 
$n$ is small, typically of the order of $10 - 40$ for small
RNA molecules. In this case the knowledge of the free-energy per monomer,
$w$, is not sufficient, and a more detailed expression for $W$
is necessary to avoid inconsistencies.

As a starting point of the analysis, we consider a polymer made of
 $N$ indentical monomers whose endpoints
are denoted by $u_i \, , \, i= 1 \cdots N$ with 
$u_0=0$.
The Hamiltonian of the chain is the sum of pairwise
interactions $\f(u_i-u_{i-1})$ and the free energy
reads, for $x=u_N$:
\beq
e^{-\b W(x,n)} = \ell_0^{-N+1} 
\int du_1 \cdots du_{N-1} e^{-\b \sum_i \f(u_i-u_{i-1})} \ ,
\eeq
where $\ell_0$ is a reference microscopic length scale.
From the above relation it follows the Chapman-Kolmogorov
equation
\beq\label{Wrec}
e^{-\b W(x,n+m)} = \ell_0^{-1} \int dy e^{-\b W(y,n)-\b W(x-y,m)} \ .
\eeq
We consider first for simplicity the Gaussian model,
$\f(x) = \frac12 k^m x^2$. Then it is easy to show that
\beq \label{casegau}
W(x,n) = \frac{k^m}{2 n} x^2 -\frac{k_BT}2 \log\left[ {\frac{k \ell_o^2}{2\pi 
k_BT\; n}}\right] 
\ . 
\eeq
In the limit of large polymers one obtains the free-energy of a monomer 
of extension $l$ through
\beq
w(l) = \lim _{n\to \infty} \frac 1n\ W(x=l\,n,n) = \varphi(l)
\eeq
as expected and consistently with the discussion of section~\ref{thermo}.
The logarithmic term in (\ref{casegau}) does contribute neither to $w$, nor
to the Langevin equation for $x$. However it does  contribute to 
the rate to close
a base-pair, see Eq.~(\ref{ratesZ}), and should be
taken into account in order to recover the correct rates.
An example of the effect of this term is obtained by computing
the equilibrium probability of $n$. Consider the (unrealistic)
case of a homopolymer, $G(n;B) = n g_0$, subject to a constant 
force and using a Gaussian model for the open part of the 
molecule; then
\beq\begin{split}
P_{eq}(n) &= \frac{1}{Z} \int dx e^{- n \b g_0 -\b W(x, 2 n) + \b f x} \\
& = \frac{1}{Z'} e^{- n \b g_0 + \frac{n}k f^2} \ .
\end{split}
\eeq
Therefore $P_{eq}(n)$ is a pure exponential, while if the
correction were neglected one would have obtained a wrong
behavior at small $n$.

For a generic model of $\f(x)$, one cannot compute $W(x,n)$.
Still we found that for our purposes ($n \gtrsim 40$) a
consistent approximation is obtained by keeping only the
first correction to the $n\to\io$ result, \ie by
defining:
\beq\label{freefinal}
e^{-\b W(x,n)} = e^{-\b n w(x/n)} \sqrt{\frac{\b k(x/n) \ell_o^2}
{2\pi n}} \ ,
\eeq
where $k(l) = w''(l)$.
One can check that this expression satisfies Eq.~(\ref{Wrec})
with corrections in the exponent of 
$O(1)$, while the terms $O(\log n+\log m)$
are taken into account.
Within this approximation, 
the error in $\log r_c(x,n)$ in Eq.~(\ref{ratesZ}) is $O(1/n^2)$
while if the first corrections are neglected it is $O(1/n)$.

In the following we will make use of the definition (\ref{freefinal})
unless otherwise stated. We will discuss an example where the
effects of neglecting the corrections is clearly observable.

\subsection{Details of the numerical simulations}

We performed numerical simulations of the molecular
constructions depicted in figure~\ref{fig:setup}, with
the following specifications:
\begin{itemize}
\item 
The total free energies of the two setups
are given by Eq.~(\ref{fafb}) plus the term $G(n;B)$.
\item 
The free energy of each polymer includes
the saddle-point corrections, \ie it is given by Eq.~(\ref{freefinal}).
The relation $l(f)$, see section~\ref{thermo}, is numerically
inverted to obtain $w(l)$ and $k(l)$ that enter in Eq.~(\ref{freefinal}).
\item 
For the single-stranded DNA we used the MFJC model,
Eq.~(\ref{FJC}), with $d=0.56$ nm, $b=1.4$ nm and
$\g_{ss}=800$ pN.
\item
For the double-stranded DNA we used the WLC model in Eq.~(\ref{DS}), 
with a small
regularization term to avoid a divergence for $f\to 0$, which
is however irrelevant for values of forces to be discussed in the
following, and with $A=48$ nm, $L=0.34$ nm and $\g_{ds}=$1000 pN.
\item
Unless otherwise stated, the double-stranded DNA linker is made by 
$N_{ds}=3120$~bps, while the two single-stranded linkers are made
by $N_{ss}=40+n$~bases each, where $n$ is the number of open DNA bases
(in other words we included on each side a 40 bases single-stranded linker).
\item
We worked at fixed temperature $k_B T = 4$ pN nm, corresponding
to $T = 16.7 \  ^\circ$C.
\item
We used the dynamical
equations for the polymers 
defined above, Eqs.~(\ref{text:finalchain}), within the discrete procedure
illustrated in section~\ref{sec:fork} and with
transition rates (\ref{ratesZ}) for the fork with attempt
rate $r=10^6$ Hz.
\item 
The matrices $\G$ corresponding to the setups in 
figure~\ref{fig:setup} are given in Eq.~(\ref{gammaagammab});
we used $\g^m_{ds}=\g^m_{ss}= \g'=2\cdot 10^{-8}$ pN s/nm.
We used a value $\g=1.67\cdot 10^{-5}$ pN s/nm for the viscosity
of the beads.
\item 
The time step was fixed to $\D t = 10^{-8}$ s; this value
ensures a correct integration of the equation of motion in all
the regimes discussed below. Even if in some cases a larger integration
step could be used, we decided to keep it fixed in order to be sure 
that discretization biases are not present.
\end{itemize}
The values of the spring constants $k_1$ and $k_2$ and of the force
$f$ in Eq.~(\ref{fafb}) varied in different simulation runs and 
will be specified later.

The program we used for the numerical simulations can be downloaded
from \url{http://www.lpt.ens.fr/~zamponi}. A user-friendly
version will be made available as soon as possible.

\subsection{Limits of validity of the dynamical model}
\label{sec:limits}

Our model of the polymer dynamics suffers from two 
main limitations.

First we keep only one collective coordinates for each polymer (its
extension) associated to the longest relaxation mode. Faster modes
are discarded. The approximation is justified provided there is no other
mode slower than the typical sojourn time on a base-pair. From the
discussion of Section \ref{ts} the number of unzipped base-pairs, $n$, 
cannot be well above a thousand.

Another upper limit on $n$ comes from the assumption that the force is
uniform along the polymer. In principle the force is a function of the
time $t$ and the location $y$ along the polymer, which obeys a diffusion
equation with microscopic diffusion coefficient $D_{ss}^m\simeq (x^m_{ss})^2/\t^m_{ss}$ 
where
$x^m_{ss}$ is the length of a monomer and $\t^m_{ss}=\g^m_{ss}/k^m_{ss}$ 
its relaxation time.
Assume that, at time 0, a base-pair closes and
the polymer is stretched at the extremity $x=0$ by $x^m_{ss}$. Then the
force, initially equal to $f(x,t=0)=k^m_{ss}\; x^m_{ss} \delta(x)$ 
will decay following
the Gaussian diffusion kernel. At time $t$ the force density at the extremity
is $f(x,t)=k^m_{ss}\;x^m_{ss}/\sqrt{2\pi D_{ss}^m\; t}$. The relaxation is
over when this force excess is of the same order of magnitude as the typical
thermal fluctuations $\delta f$ calculated in (\ref{fluc1}), that is,
for times
\beq 
t > n\ \frac {k^m_{ss} \; (x^m_{ss})^2}{2\pi \; k_BT}  \; \tau^m_{ss} \simeq 
2\; 10^{-10}\; n \ \hbox{\rm ps} \ .
\eeq
When $n\sim 1000$ the corresponding relaxation time is of the order of the 
sojourn time on a base. 

In conclusion our dynamical model is adapted to ssDNA polymers whose length
ranges from a few hundreds to a few thousands bases.
Shorter polymers can be considered at equilibrium, while longer polymers
cannot be modelled without taking account the space-dependence of forces.
A simple way to tackle this difficulty consists in arbitrarily cutting 
long polymers in 1000-base long segments, each modelled as above. This 
procedure will be followed in Section \ref{sec-corr}.

\section{Unzipping at fixed force}
\subsection{Quasi-equilibrium unzipping}
\label{sec:quasi}

Before turning to the more interesting case of
out-of-equilibrium unzipping, we focus on the case
of a small molecule which is subject to a constant force
close to the critical force. In this situation the molecule
is able to visit all the possible configurations.

\begin{figure}[t]
\includegraphics[width=9cm]{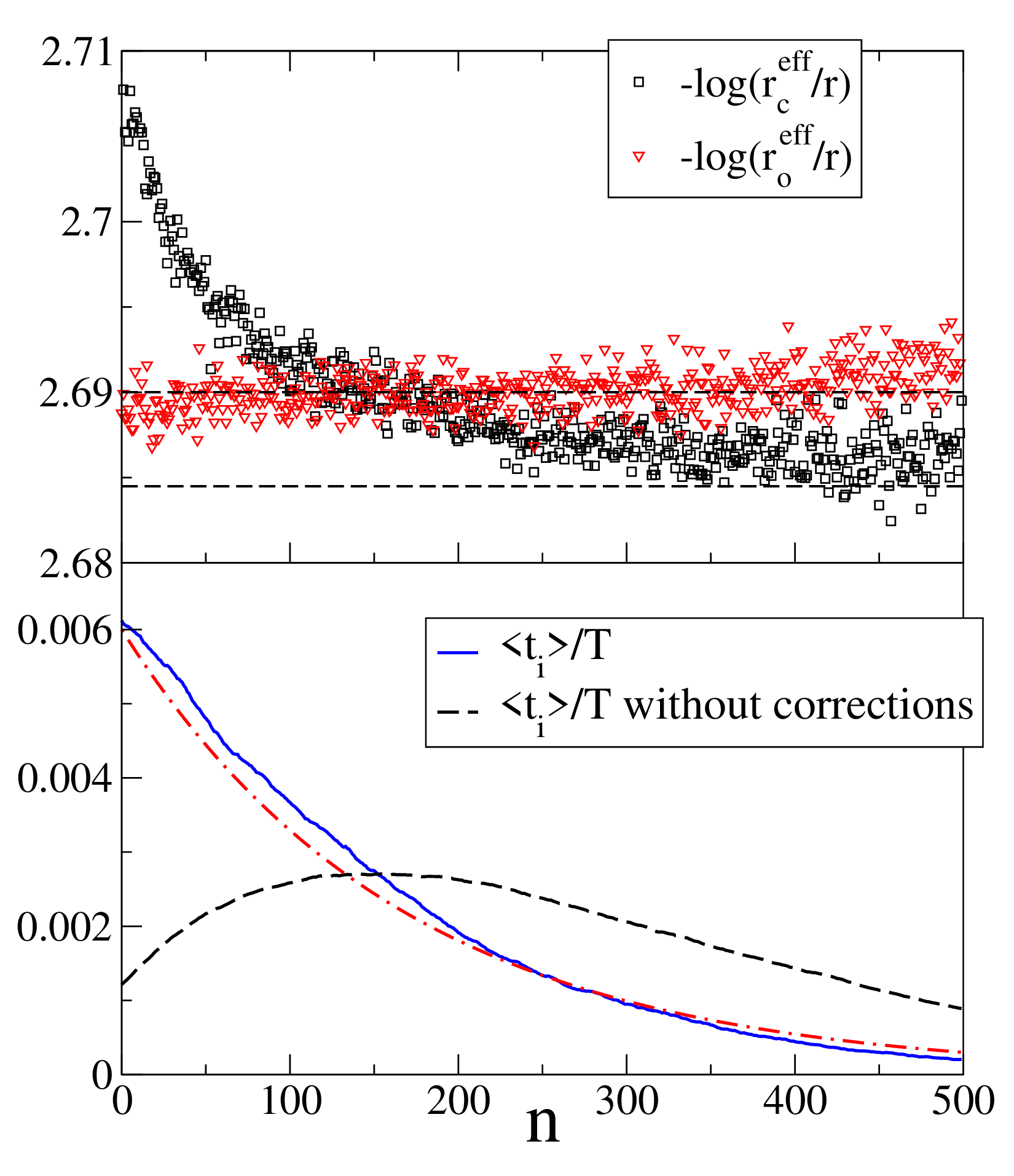}
\caption{
{\it Bottom:} Average fraction of 
time spent on each base. 
The full (blue) curve corresponds
to Eq.~(\ref{freefinal}) while the dashed (black) curve
corresponds Eq.~(\ref{freefinal}) without the saddle-point 
corrections (the square-root term).
The dot-dashed (red) line is 
$P_{eq}(n) \propto \exp [ - n \, \D g ]$
with $\D g = 0.006$.\\
{\it Top:}
Effective rates (squares and triangles)
estimated from
the maximization of the probability in 
Eq.~(\ref{Prateeff}) ($r=10^6$ Hz) without
saddle point corrections (full curve of the
lower panel).
The dashed lines are the asymptotic values of
the rates, see text.
We do not report the rates corresponding to
the full Eq.~(\ref{freefinal}) since they are essentially
independent of $n$.
}
\label{fig:rates}
\end{figure}

We performed a set of numerical simulations at constant
force $\bar f = 16.45$ pN, with the setup described in
figure~\ref{fig:setup}B. The DNA molecule is a uniform segment of
$N=500$ base-pairs, with pairing free energy $G(n;B)=  n g_0$ and
$g_0 = 2.69 \, k_B T$. The entropic free energy
per base of the two open single strands is
$2 g_{ss}(\bar f) = 2.684 \, k_B T$, therefore the infinite
molecule would stay close; we are sligthly below the critical
force. To the right and left 
open portions of the molecule, two single-stranded
DNA linkers of $N^0_{ss}=40$ bases each are attached; therefore
the total length of the single-stranded linkers is
$N_{ss}=N^0_{ss} + n$, where $n$ is as usual the number of
open base-pairs.
The leftmost linker is a double-stranded DNA of $N_{ds}=3120$ base-pairs, whose presence is however irrelevant for the scope of this
section.
The total length of the simulation was
of $T=7200$ s, \ie of two hours.

\subsubsection{A test of the model}

The average fraction of 
time spent on each base, corresponding to the equilibrium
probability distribution $P_{eq}(n)$, is reported in the lower
panel of figure~\ref{fig:rates}. We expect that in the large
$n$ limit, $P_{eq}(n) \sim \exp [ -n (g_0 - 2 g_{ss}(f)) ] = 
\exp [ - n \, \D g ]$, with $\D g \sim 0.006$. This is expected
to break down when $N_{ss}$
is so small that the second-order
corrections to the saddle-point in Eq.~(\ref{freefinal})
become important. 
As it can be seen in figure~\ref{fig:rates}, the exponential
form correctly describe the data.

We performed additional simulations in which the square-root
term in Eq.~(\ref{freefinal}) was removed. As one can see, in
this case the small $n$ deviations are much more pronounced.
It is worth to note that for a non-Gaussian polymer one expects
a deviation from the exponential form at small enough $n$.
However, this analysis shows that taking into account the small
$n$ corrections to $W(x,n)$ systematically reduces this effect.
Estimating its real order of magnitude therefore requires an exact
expression of $W(x,n)$, which could be in principle obtained
from the recurrence equation (\ref{Wrec}). However this is a
complicated numerical task that goes beyond the scope of this
article. What we want to stress here is that the inclusion of
the square-root term in Eq.~(\ref{freefinal}) 
gives significant differences when $n \lesssim 200$ and should
therefore be included if one wants to analyze the unzipping
of small molecules.

\subsubsection{Effective dynamics of the fork}

In a situation where the linkers are short, such that their
relaxation time is faster than the mean time spent on a base,
the linkers are able to reach equilibrium before $n$ changes.
Therefore one might hope to define an {\it effective dynamics}
for the fork, where $n$ changes according to effective rates
that depend on the variation in free energy of the setup
on closing or opening a base.

To this aim we considered the model for the fork dynamics 
described in section~\ref{direct}, but assuming
$n$-dependent opening and closing rates. 
Within this model,
the probability of a trajectory
of the fork is a function of the number of upward ($u_n$)/downward ($d_n$) 
jumps, and the time spent on base $n$, $t_n$:
\beq\label{Prateeff}
\begin{split}
P_{eff}[n(t)] &= \prod_{n=1}^N (r^{eff}_c(n) \D t)^{d_n} 
(r^{eff}_o(n) \D t)^{u_n} \\ 
&\times (1 - \D t(r^{eff}_c(n)+r^{eff}_o(n)))^{t_n}
\end{split}\eeq
Given the values of $u_n,d_n,t_n$ measured along our trajectory
of duration $T$, we can infer the effective rates by maximizing
the above probability. Assuming $r^{eff} \D t \ll 1$ we obtain
\beq
r_c^{eff}(n) = \frac{d_n}{t_n} \ , \hskip1cm
r_o^{eff}(n) = \frac{u_n}{t_n} \ ,
\eeq
as estimates for the effective rates. For
the full expression (\ref{freefinal}), the rates are almost
independent of $n$; on the other hand, if the first order
correction are neglected, one obtains $n$-dependent rates,
consistently with the observation that $P_{eq}(n)$ is not
exponential. These are reported in the
upper panel of figure \ref{fig:rates}.
In both cases, the rates are consistent with the
detailed balance condition 
$r^{eff}_c(n) P_{eq}(n) = r_o^{eff}(n-1) P_{eq}(n-1)$.

\subsection{Out-of-equilibrium opening} 

For long molecules the barrier
between the closed and open states may become very large e.g.  $\sim
3000$ k$_B$T for the 50000 bases $\lambda$--DNA at the critical force
$f_c=15.5$ pN \cite{Coc4}. The time necessary to cross this barrier is
huge, and full opening of the molecule never happens during
experiments.  To open a finite fraction of the molecule the force has
to be chosen to be larger than its critical value. The opening can 
then be modeled as a transient random walk, characterized by pauses at local
minima of the free energy and rapid jumps in between \cite{Dan03}.  

\subsubsection{Analytical calculation of the average time spent by the fork on a base}
\label{sec:caltheo}

Consider first the case of a fixed force acting on the fork while
all the other components are at equilibrium as in Section
\ref{direct}. In the transient random walk the opening fork
spends a finite time around position $n$ before escaping away and never
coming back again in $n$.  The number $u_n$ of opening transitions
$n\to n+1$, is stochastic and varies from experiment to experiment,
and base to base. The total number of times the fork visits base-pair
$n$ before escaping is given by the sum of the number $u_n$ of
transitions from $n-1$ to $n$ and of the number $u_{n+1}-1$
 of transitions from $n+1$ to $n$.  The average time spent in $n$ is 
therefore
\begin{equation} t_n= \frac{\langle u_n\rangle+\langle u_{n+1}\rangle-1}
{r_o+r_c(n)}
\label{tn}
\end{equation} where $1/(r_o+r_c(n))$ is the average time spent in $n$
before each opening or closing step.  Let us introduce
the probability
$E_{n+1}^n$ of never reaching back position $n$
starting from position $n+1$.  The probability $P$ of the number
$u_n$ of opening transitions $n\to n+1$ during a single unzipping
simply reads
\begin{equation} P(u_n)= \left(1-E_{n+1}^n\right)^{u_n-1}\;E_{n+1}^n
\label{defrho1}
\end{equation} From equation~(\ref{defrho1}) we have that the average
number of openings of bp $n$ is
\begin{equation} \langle u_n \rangle = \sum _{u_n\ge 1} P(u_n) \; u_n 
=\frac{1}{E_{n+1}^n} \ .
\label{un}
\end{equation} 

We are thus left with the calculation of $E_{n+1}^n$.
For infinite force $E_{n+1}^n=1$ since the fork never moves backward.  
For finite force we write a recursive equation for
the probability $E^n_m$ that the fork never comes back to base $n$
starting from base $m \;(\ge n+1)$
\begin{equation} 
E_m^n=q_m E_{m-1}^n+(1-q_m)E_{m+1}^n \ .
\label{e}
\end{equation} 
where
\begin{equation}
q_n=\frac{e^{g_{ss}(f)}}{e^{g_{ss}(f)}+e^{g_0(b_n,b_{n+1})}}
\end{equation} 
is the probability to close base $n$ and $1-q_n$ is the
probability to open it at each step. Note that for forces larger than
the critical force, we have $q_n < \frac{1}{2}$: the random walk is
submitted to a forward drift and is transient.  
The boundary conditions for equation (\ref{e}) are
$E_n^{n}=0$ and $E_m^{n}=1$ for $m\to\infty$.   

For an homogeneous
sequence the escape probability is $E=(1-2q)/(1-q)$. For an
heterogeneous sequence by defining $\rho^n_m=\frac{E^n_m}{E^n_{m+1}}$
we obtain the Riccati recursion relation
\begin{equation} 
\rho^{n}_n=0\,; \ 
\rho^{n}_{m+1}=\frac{1-q_{m+1}}{1-q_{m+1}\;\rho^{n}_m}
\hspace{1cm} {\mbox {for}\ n\geq m} \ .
\label{ric}
\end{equation} 
Equation (\ref{ric}) can be solved numerically for a
given sequence. The escape probability starting from $n+1$ is
then
\begin{equation} E _{n+1}^n = \prod_{m\geq n+1} {\rho_{m}^{n}} \ .
\label{pj}
\end{equation} and the average time spent in the base $n$ is then
obtained from (\ref{un}) and (\ref{tn}).

\begin{figure}[t]
\includegraphics[width=.45\textwidth]{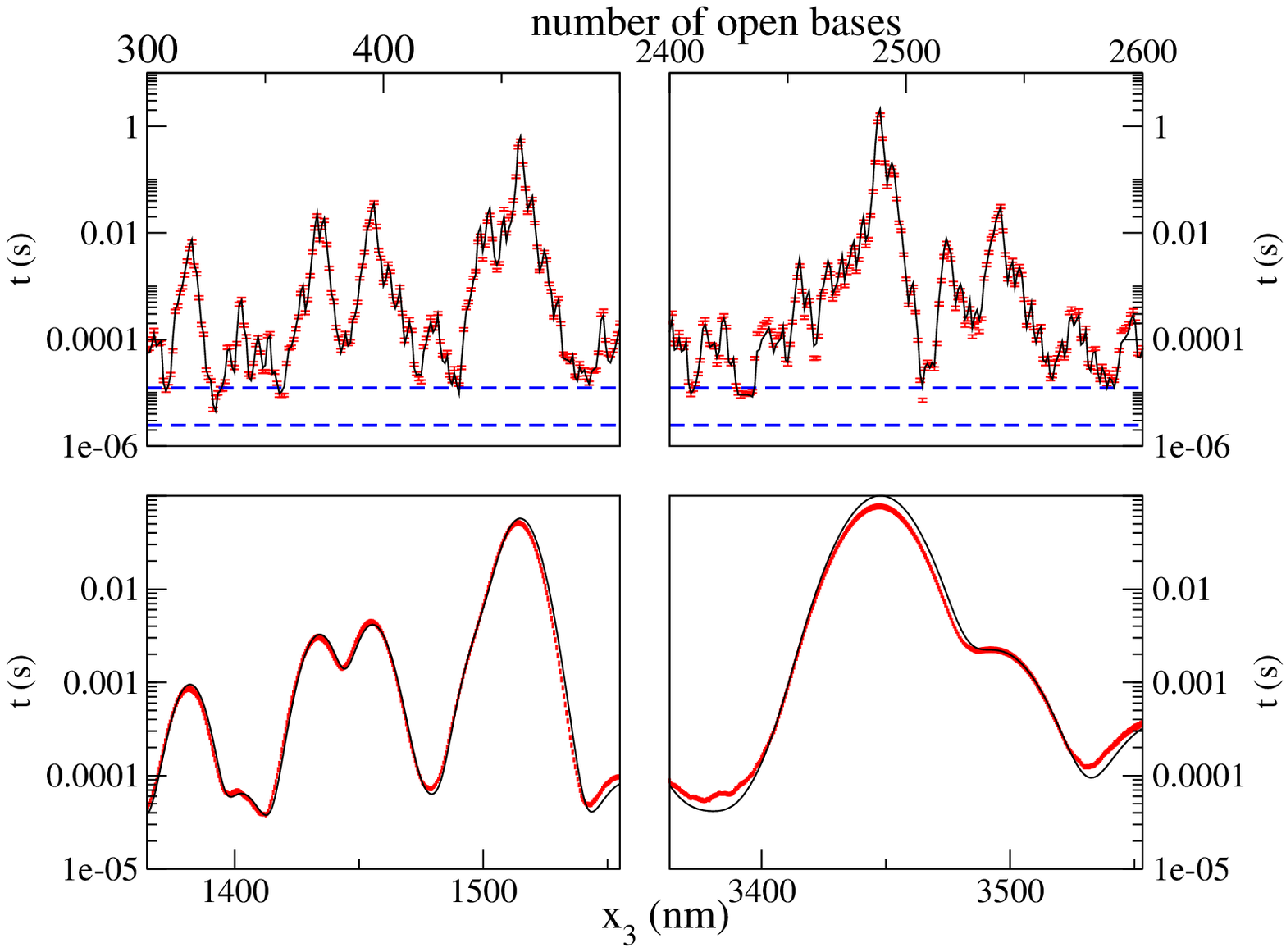}
\caption{{\it Top:} Average time spent by the fork on position $n$. \\
{\it Bottom:} Time
spent by the whole setup at extension comprised between $x_3$ and 
$x_3+\Delta x$, with $\Delta x=0.5$ nm. \\
The black line in both figures represent the theoretical 
predictions from Section \ref{sec:caltheo}. 
Red points are the results from the simulation. Standard deviations are
represented by error bars in the top panels, and by the thickness of 
the red curves in the bottom panels.}
\label{fig:timeout}
\end{figure}

\subsubsection{Results from the dynamical model}

To check whether these theoretical predictions are affected by 
dynamical fluctuations of the bead, linkers, and unzipped strands we have 
carried out simulations with the model of Section \ref{sec:model}. We have
carried out 160 unzippings of the $\lambda$-phage sequence
at the force of $17$~pN for $T=100 $ seconds (physical time), 
with the same molecular construct of section~\ref{sec:quasi} 
($N_{ds}=3120$  base-pairs dsDNA linkers
on a side plus $N^0_{ss}=40$ bases of ssDNA linker at each side 
of the DNA to be open).
For such a construct the equilibrium extension
of the polymers for $n$ open base-pairs is $2N_{ss} l_{ss}+N_{ds} l_{ds}$
where $l_{ds}=0.3337$ nm,and $l_{ss}= 0.4758$ nm, and $N_{ss}=N^0_{ss}+n$.
The stiffness of the polymers is 
$1/k_{eff}=N_{ss}/k^m_{ss}+N_{ds}/k^m_{ds}$
 with $k^m_{ss}=160.5$ pN/nm and $k^m_{ds}=1450$ pN/nm.
The relaxation times of the polymers are
of the order of 0.1 ms for about 400 unzipped bases and 1 ms for about
2500 open bases, and are larger than the characteristic times 
to open a weak base, of about $2\;10^{-6}$~s, and a strong base, of 
about $10^{-5}$~s.

We plot in Figure \ref{fig:timeout} the average time spent by the fork 
at location $n$ for two portions of the sequence, corresponding to 
about 400 and 2500 open base-pairs. 
The agreement between  the theoretical and numerical estimates of the times
is excellent, meaning that the fluctuations of extensions of the polymers
and the dynamics of the bead induce negligible changes on the rates of
opening and closing, as seen close to the critical force in Section
\ref{sec:quasi}. 

As experiments do not give direct acces to the time spent by the fork at 
location $n$ we show in Figure \ref{fig:timeout} (bottom) 
the time $t(x_3)$ spent by the unzipped ssDNA 
between extensions $x_3$ and $x_3+dx$. These times are compared to
their values assuming that the positions $x_3$ of the beads are 
randomly drawn from the equilibrium measure,
\beq
t(x_3) = \sum _n t_n \ P(x_3|n) \ ,
\label{eq:convol}
\eeq
where $t_n$ is calculated from (\ref{tn}) and $P(x_3|n)$ is calculated from an argument similar to those used in section \ref{sec:manip} and can be written up to quadratic order around the saddle point as:
\begin{eqnarray}
P(x_3|n)&=&
\sqrt{\frac{\beta k_{eff}(f)}{2\pi}} \\
&\times&  e^{-\beta \frac{k_{eff}(f)}{2}\left(x_3-N_{ds}l_{ds}(f)-2N_{ss}l_{ss}(f)\right)^2}\nonumber\,.
\end{eqnarray}
The agreement is, again, excellent. 

Figure~\ref{fig:timeout} and equation (\ref{eq:convol}) show
that $t(x_3)$ gets contributions from the times
spent by the fork on a set of bases whose number depends on the magnitude
of the equilibrium fluctuations of the linkers. 
These equilibrium fluctuations increases with the length of ssDNA e.g. 
$\delta x_3 \simeq 5$ nm for 400 unzipped base-pairs
  and $\delta x_3 \simeq 12$ nm for 2500 unzipped bases.
Therefore, as the number $n$ of unzipped base-pairs increase, the characteristic
curve of $t(x_3)$ gets more and more convoluted (compare left-bottom and
right-bottom panels in \ref{fig:timeout}). 

\begin{figure}[t]
\vskip 1.6cm
\includegraphics[width=.45\textwidth]{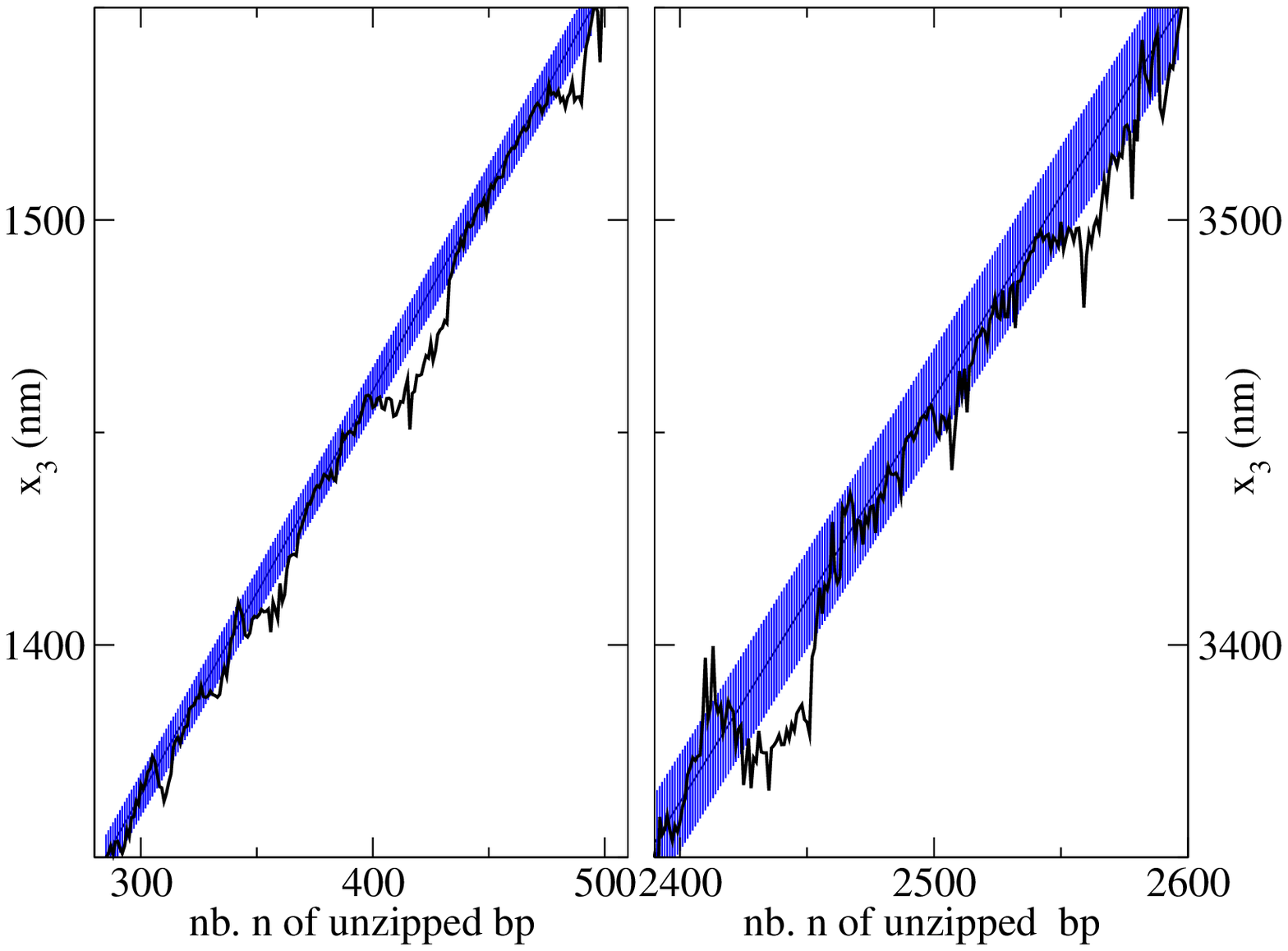}\\
\vskip 1cm
\caption{Total extension $x_3$ of the setup in figure~\ref{fig:setup}B
at fixed number $n$ of unzipped bases 
for a single unzipping (black line). If the fork visits twice or more the
same base $n$ we plot the average of the extension values. The gray strip 
represents the average value at equibrium, $x_3^{eq}(n)$, and the standard
deviation around its value at equilibrium.}
\label{fig:xeq}
\end{figure}

In Figure~\ref{fig:xeq} we compare the value of the
ssDNA extension from one unzipping, $x_3$, 
to its average value at equilibrium,
$x^{eq}_3$, as a function of the number of unzipped base-pairs $n$.
The fluctuations in the extension are compatible with the
equilibrium deviations. Again no clear out-of equilibrium effect is observed.
The reason is that, even if the single strand is not relaxed in the 
opening time of a base, the fork goes back and forward
around a given location before moving away. 
Therefore the quantities we have measured are averaged
on the number of times a base-pair is opened
and are close to their mean value even in a single unzipping. 
This can be deduced from 
figure~\ref{fig:timeout} by
comparing the total time spent on a base (points)
with the time to open a base (dashed lines)

\section{Unzipping at fixed extremities}
\subsection{Correlation functions}
\label{sec-corr}
\begin{figure}[t]
\includegraphics[width=9cm]{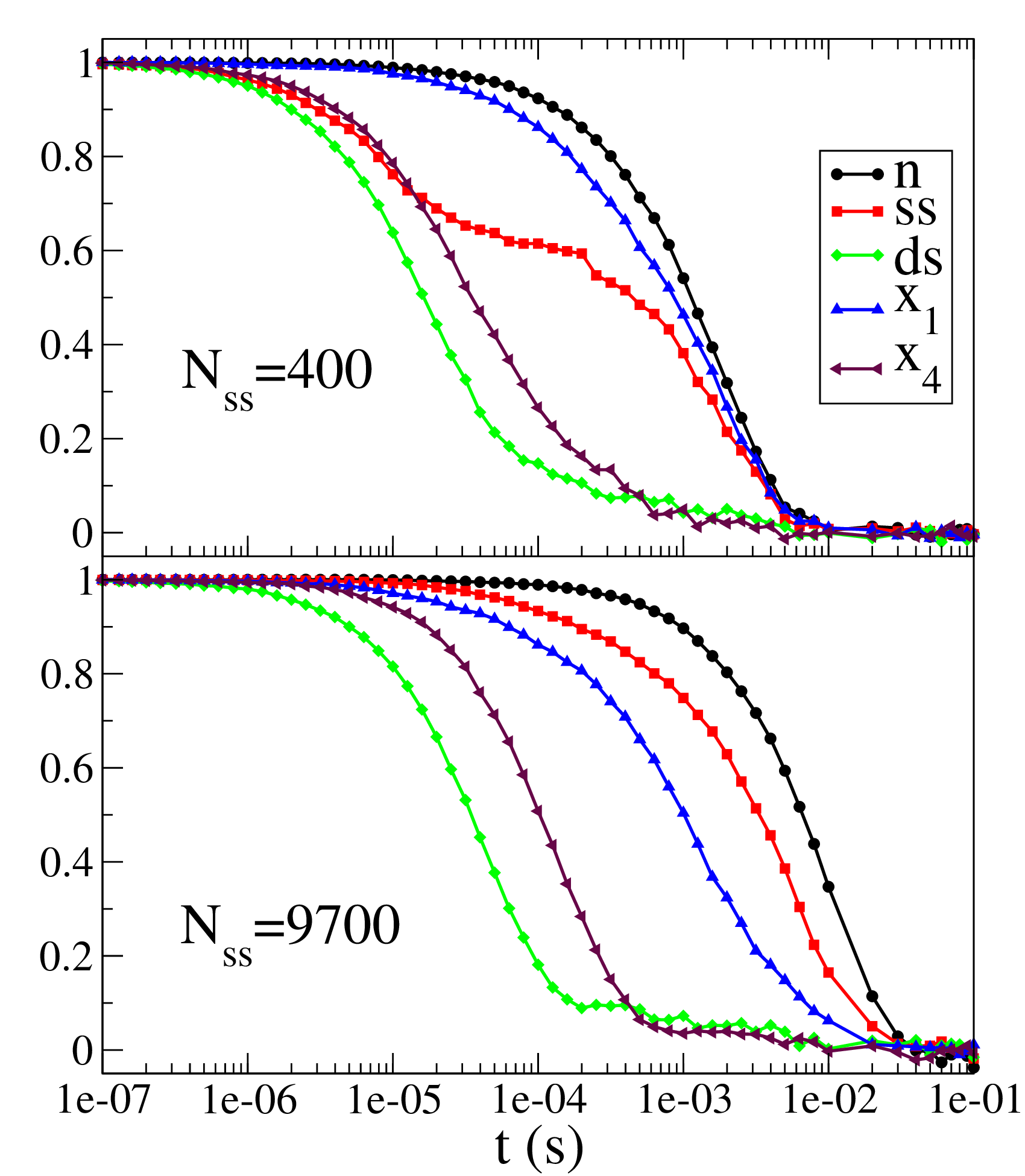}
\caption{
Correlation functions for the setup in
figure~\ref{fig:setup}A at two different values of
the number of open bases, $N_{ss}=40 +n$.
}
\label{fig:corren}
\end{figure}

One of the main advantages of considering the dynamics of the linkers and of the beads is that it allows us to compute autocorrelation functions and to explore the interaction between the different parts of the setup, a task which would be impossible from \emph{a priori} calculations.\\
We have performed a few simulation with the setup shown in figure \ref{fig:setup}A where the spring constant of the first optical trap of extension $x_1$ is 0.1 pN/nm and the second ($x_4$) has stiffness 0.512 pN/nm. The molecule in the fork is uniform with $g_0 = 2.69 k_BT$. 
The only parameter that is varied across simulations is the distance between the optical traps and thus the typical number of open bases.\\
In figure \ref{fig:corren} we show two typical cases. What is evident is that the single strand has two time scales: one which is proper to the fluctuations at $n$ fixed and another which is of the same order of magnitude as the correlation time of the fork. As the number of open bases grows the fast time scale also grows until 
it becomes impossible to distinguish the two.\\
As remarked in section~\ref{sec:limits}, our model cannot in principle 
be used when the linkers are made of $n \gtrsim 1000$ monomers. 
To check for the importance of force propagation effects, we ran a simulation
for $N_{ss}=9700$ (bottom panel of figure~\ref{fig:corren})
where we cut each linker in 9 subunits of 1000 bases each, plus a final unit which is
connected to the opening fork. 
Overall, the correlation functions are not much affected by this modification and in 
particular the correlation times are unaffected within numerical errors.
The main effect of cutting the long linkers is that the correlation function of the
linker becomes more strectched (i.e. if they are fit with $\exp[-(t/\t)^{\b_s}]$, the
exponent $\b_s$ is sligthly smaller). This is to be expected since by cutting the
polymer we include more relaxation modes, each with its relaxation time.
A wider distribution of relaxation times implies a smaller exponent $\b_s$.

In table \ref{tab:confronto} we compare the results of the numerical simulation with 
the predictions of section \ref{sec:manip} which do not take into account the interactions between different parts of the setup. While the simulated result for the single-stranded and the double-stranded DNA are not too far off from the prediction, the two springs show a much greater deviation from the theoretical estimates. This prompted us to analyse further the relationship between the fork and the bead position as will be discussed later.\\
The potential acting on the fork position, in the case of an uniform molecule is dictated by the stiffness of the rest of the setup only as seen in section \ref{sec:manip}. That is to say that $n$ experiences an harmonic potential with spring constant proportional to $k_{eff}$; this in turn predicts correlation times that are proportional to $\frac1{k_{eff}}$ which has a linear dependence on $n$. This behavior is in very good agreement with the data that has been extracted from numerical simulations.
\begin{table}
\begin{center}
\begin{tabular}{|l|l|l|}
 \hline& Theoretical (s) & Numerical (s) \\ \hline
Single strand & $4.83\cdot 10^{-11}N_{ss}^2$ & $5.4\cdot 10^{-11} N_{ss}^2$\\ \hline
Double strand & $4.96\cdot10^{-5}$ & $\sim 3\cdot 10^{-5}$ \\ \hline
Spring $x_1$ & $1.67 \cdot 10^{-4}$ & $\sim 1.5\cdot 10^{-3}$ \\ \hline
Spring $x_4$ & $3.26 \cdot 10^{-4}$ & $\sim 7 \cdot10^{-5}$ \\ \hline
Fork $N_{ss}$ & $\propto 14.2 + 0.013 N_{ss}$& $1.3 \cdot 10^{-3}+8.4 \cdot 10^{-7} N_{ss}$  \\ \hline
\end{tabular}
\end{center}
\caption{Comparison between the correlation times of the setup in figure~\ref{fig:setup}A
as computed for an isolated element 
and the result of a complete numerical simulation.
In the case of the fork, we reported as theoretical value $1/k_{eff}$, that must be
multiplied by a viscosity to obtain the relaxation time; it turns out that a viscosity
$\sim 8 \cdot 10^{-5}$ pN s/nm matches the theoretical and numerical results.
}\label{tab:confronto}
\end{table}

\begin{figure}[t]
\includegraphics[width=9cm]{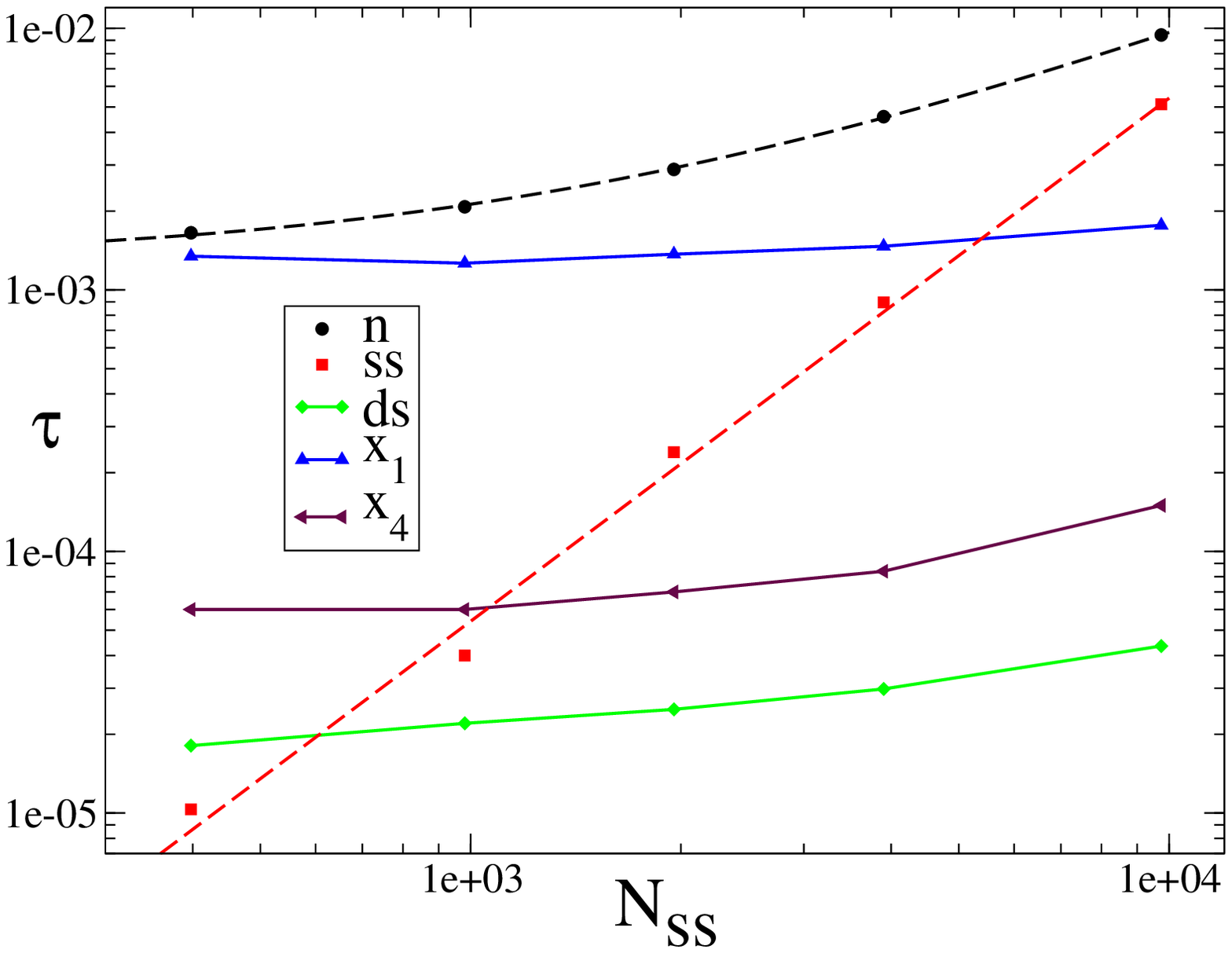}
\caption{
Relaxation times of the correlation functions
in figure \ref{fig:corren} as a function
of the number of open bases. In the case of the single strand (ss), only the fast relaxation time is plotted. 
For the fork and the single strand, 
dashed lines indicate a fit to $\t_n = A + B N_{ss}$ 
(with $A=1.3 \cdot 10^{-3}$ and $B=8.4 \cdot 10^{-7}$)  
and $\t_{ss} = C N_{ss}^2$ (with $C=5.4 \cdot 10^{-11}$ s). 
For the others, full lines are guides to the eye.}
\label{fig:times}
\end{figure}

\begin{figure}[htbp]
\includegraphics[width=9cm]{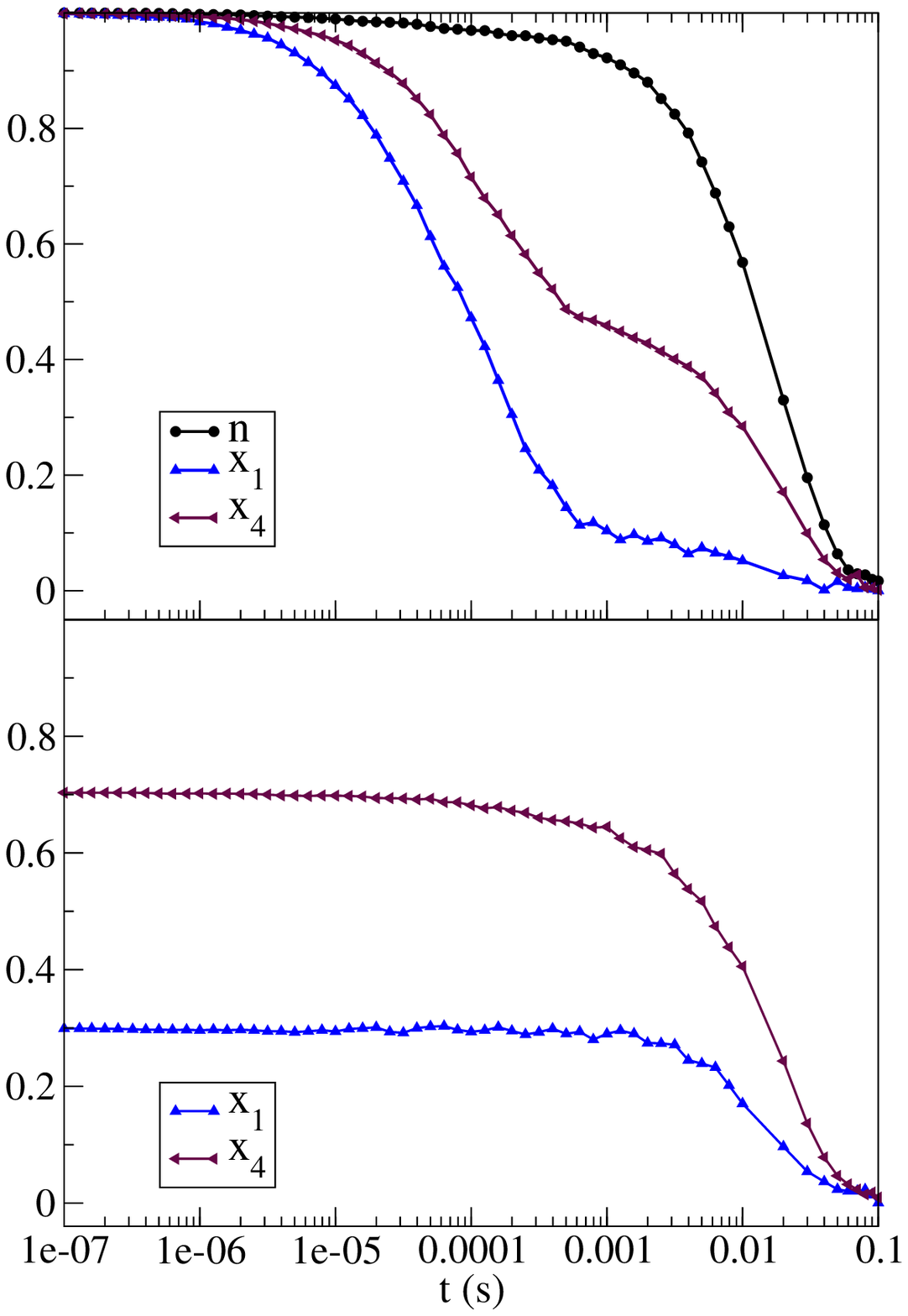}
\caption{\emph{Top}: Autocorrelation functions for the setup in
figure~\ref{fig:setup}A when the molecule to unzip is a block-copolymer composed of alternating stretches of ten strong pairs and ten weak pairs. This way the fork correlation time is greatly increased allowing us to view effects on the two traps of different optical stiffness.\\
\emph{Botttom}: Correlation functions between one of the two beads and the number of open base-pairs. Values have been normalised so that the value at zero time difference is $\rho=\langle x_i n\rangle/\sqrt{\langle x_i^2\rangle\langle n^2\rangle}$}
\label{fig:springs}
\end{figure}

\subsection{Mutual information between bead position and fork location}
\label{info}

Figure \ref{fig:springs} shows the dynamical correlations
of the fork and beads positions. The two beads have
different correlations functions due to the difference in their
stiffnesses: $k=0.5$ pN/nm for bead 1 and $k=0.1$ pN/nm for bead
2. After an initial decay (taking place over a time
proportional to $1/k$ from Section \ref{IIC3})
the bead correlations exhibit a quasi-plateau
behavior whose height is roughly proportional to $1/k$. The plateau
reflects the correlation between the motion of the bead and the one 
of the fork on time-scales of the order of the equilibration time of 
the fork. It appears that soft beads allows one to track the location of the
fork better than stiffer beads.\\
In the following we will give a closer look to the dependence of this correlations to the optical trap stiffness, to do so we construct a setup as in figure~\ref{fig:setup}A, but where the stiffness of the optical trap on the left is kept constant at 0.512 pN/nm while the stiffness of the one the right is varied across two orders of magnitude\footnote{The attentive reader might have noticed we changed the stiffness of the right bead compared to what it was in the previous section, the rationale behind this choice is to keep its value at the center of the range in which we will vary the other.}.

To give quantitative support to this statement we
define the mutual information $I$ between the position of the bead
in the optical trap, $x_4$, and the number of open base-pairs, $n$:
\begin{equation}
 I(x_4,n)=\sum_n\int dx_4 P(x_4,n)
 \log\left(\frac{P(x_4,n)}{P(x_4)P(n)}\right)\,,
\end{equation}
where $P(x_4,n)$ is the joint probability density for the bead to be
at position $x_4$ while there are $n$ open base-pairs; $P(n)$ and
$P(x_4)$ are the two marginals. Note that the definition of mutual
information does not suffer from the problems which arise with entropy when we switch
between a continuous and a discrete definition, that is to say that
binning with sufficiently small bins does not change the mutual
information.

$I$ can be easily computed by keeping track
of the times passed at a given bead position and given number of open
bases during a run of the simulation. As stressed before the fact that
the $x_4$ coordinate must be binned has negligible effects on the
computation of entropy. For very large stiffnesses the amplitude of 
the oscillations of the bead can become
very small, and thus a lack of sensitivity
in the measure of the position of the bead could become an
issue. Fortunately current state of the art in optical trap  
cannot attain stiffnesses larger than, say,
1 pN/nm with micrometer beads \cite{Mang07}. 
In this regime the fluctuations of
the bead are dominated by the stiffness of the trap and thus we can
say that $\langle\delta x_4^2\rangle \sim (\beta
k_{2})^{-1}$, see Eq.~(\ref{tabinv}). 
Comparing the fluctuations of the bead
position with the sub-nanometer precision $\Delta$ over its location
yields
\begin{equation}
 \frac{\sqrt{\langle\delta x_4^2\rangle} }{\Delta}\simeq 10 -50 \,,
\end{equation}
which is much larger than unity.

\begin{figure}[htbp]
\includegraphics[width=0.45 \textwidth]{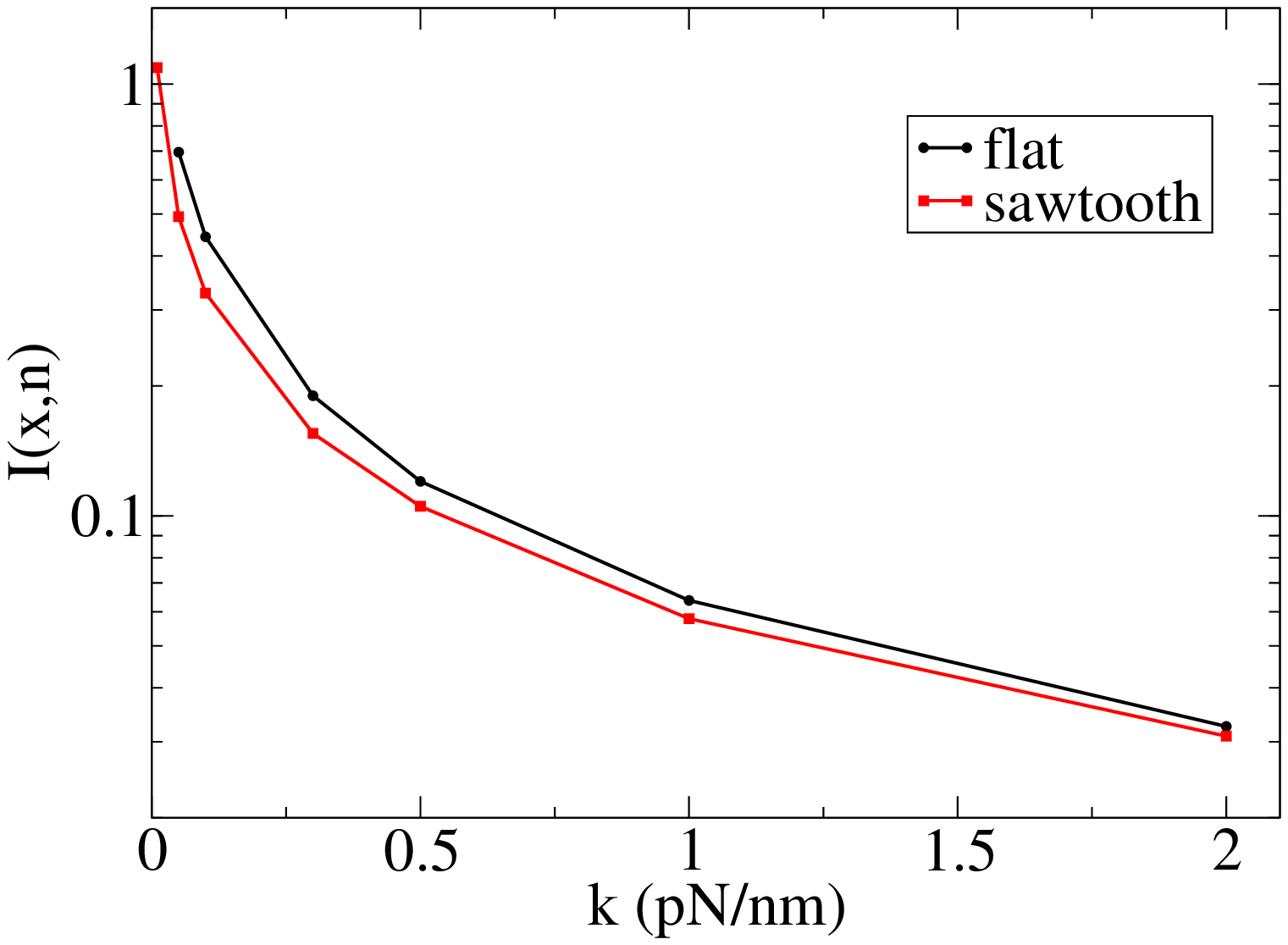}
\caption{Mutual information $I$ between $x_4$ and $n$ as a function of
the trap stiffness,  $k$. Black circles are computed on an uniform
  sequence, while red squares are measured on the sawtooth
  potential described in the caption of figure \ref{fig:springs}.}
\label{fig:entropy}
\end{figure}

Figure \ref{fig:entropy} shows that the mutual information $I$ 
depends only weakly on the sequence but depends strongly
on the stiffness $k$ of the trap. This behavior can be understood very
intuitively. Right after a base-pair  opens or closes, the whole
setup in a fixed force experiment has to give way; 
the less an element of the setup is rigid compared to the rest, the
more it will accomodate for the change in $n$.

We conclude that, in a single measurement, soft traps
give more information on the fork location than stiff traps.
However $I$ is the mutual information between the fork and bead locations
\emph{per measure}. 
As we have seen in Section \ref{sec-corr}, the correlation
times extracted from the simulations decrease with $k$ and, as $k$ grows, 
more and more uncorrelated measures can be done in the same amount of time.
It is thus expected that the information \emph{per unit of time} is 
not maximal for small values of $k$. 
In other words, stiffer traps give worse-quality
but more frequent signals on the location of the fork. Finding the optimal
value of $k$ would require a detailed analysis of the 
correlation times of the bead and of the fork.
In particular the size of the bead would affect the optimal value for
$k$ through the viscosity coefficient, but not the information per 
measure, $I$. However this dependence should not be crucial since the 
bead size cannot be varied much in 
experiments: it can be neither too small to exert a sufficient force, 
nor too large due to the size of the physical setup.

\section{Conclusion}

This paper has been devoted to the presentation of a dynamical model for
the different components of the setups used in the unzipping of single DNA
molecules under a mechanical action. Compared to previous studies our model
does not assume {\em a priori} that the polymers in the molecular construction 
are at equilibrium but takes into account their relaxation dynamics.
It is important to stress out that the dynamical description for the 
linkers and the unzipped part of DNA is coarse-grained: the basic unity 
are the polymers themselves, not the monomers they are made of.

As a consequence each polymer is associated with a unique relaxation time.
The assumption is justified as long as those times are comparable to the
typical opening or closing time of a single base-pair. Longer polymeric
chains e.g. ssDNA strands with a few thousands bases 
need to be modeled in a more detailed way; more precisely, they should be
divided into short enough segments along which the force can be considered as uniform
on the time scales associated to the fork motion. 
Although in this paper we did not observe any important force
propagation effect, these might be more important in strongly
nonequilibrium situations such as opening at constant (high) velocity.
We plan to simulate unzippings
with such molecular constructions in the next future to understand how force
propagation across the polymeric segments can affect the effective rates for
closing base-pairs in such situations. 
  
One of our results is that one has to be very careful with the expression of the
free-energies (entering the dynamical rates) for short polymers, be they 
linkers or ssDNA unzipped strands. Use of the free-energy per monomer, obtained
from force-extension measures on long molecules, as usually done in the literature
can lead to erroneous results. We have shown that finite-size corrections 
to the energetic contributions and the dynamical rates have to be taken into account.

As a main advantage the code we have developed is versatile: 
we can easily change setups, for example use fixed-force or fixed-position 
ensemble, and change the number and types of linkers and of traps for the beads.
We have found that, in fixed-force unzippings,  the opening and closing
rates for the fork are not affected by the force fluctuations coming from the
polymeric chains. For small linkers and number of unzipped base-pairs, indeed,
force fluctuations are large but fast, and are averaged out on the 
characteristic opening-closing time of a base-pair. For large linkers or 
number of unzipped bases force fluctuations are slow but small, and therefore 
do not change the dynamic of the opening fork. We have also performed  
unzipping simulations  at large forces where the opening dynamics is transient,
and found that the average time spent by the unzipped strands at a given extension 
is accurately predicted from the time spent by the fork on a base convoluted by the
equilibrium fluctuations of ssDNA. Moreover the extension between the 
extremities at a fixed number of open base-pairs in a single unzipping 
experiments is compatible with equilibrium fluctuations of ssDNA and linkers.
The program could be easily adapted to unzipping at constant velocity, 
where non-equilibrium effects are likely to be more important.

Our study suggests that one measure of the position of the bead in soft traps 
gives more information on the location of the fork than in the case of stiffer
traps. This statement is however to be considered with caution. Beads in stiffer
traps reach equilibrium on shorter time scales, and the overall
rate of information per
unit time could be higher in stiffer traps. While purely qualitative at this
stage, such a statement is relevant to the study of the inverse problem of unzipping, 
that is, inferring the sequence of the DNA molecule from the unzipping signal.
We hope that the present dynamical modeling will be useful to assess the 
rate at which information on the sequence could be acquired from mechanical single
molecule experiments.

\vskip .5cm \noindent
{\bf Acknowledgement:} We thank U. Bockelmann, I. Ciss\'e, M. Manosas, and P. Pujol
for useful discussions. 
This work has been partially funded by the PHC Galileo program for 
exchanges between France and Italy, and the 
Agence Nationale de la Recherche project ANR-06-JCJC-0051.

\appendix
\section{Langevin dynamics of coupled polymers}
\label{app:poly}

One of the simplest models of polymer dynamics is the one proposed by Rouse 
\cite{Rouse}, where the polymer is described as a chain of beads which are modeled as 
Brownian particles, linked by harmonic springs.\\
While it is true that this model is very crude because it ignores hydrodynamic 
interactions and excluded volume  effects, it has the huge advantage of being largely solvable. 
Therefore we will use it now as the basis for a few considerations 
that will be then generalized to more realistic models.

Our aim is to write a system of coupled equations for the time evolution of 
a certain number of marked points on a (hetero)polymer. One of these points will be
for instance the location of the opening fork. In the case of a double
DNA strand attached to a single strand, one point will mark the location where
the two different polymers are attached (see the examples in 
figure~\ref{fig:setup}).
Note that if the marked points we focus
on are far apart, only the slower modes of the system will be relevant, as the
fast modes describe local relaxations of the chain. Therefore in the following
we want to focus on a long wavelenght/long time effective description of the chain.

\subsection{The dynamics of a single polymer}

\subsubsection{The model and its normal modes}

As the simplest case we consider a polymer 
composed of $N$ identical springs, each with an identical 
link at one end. The first is connected to a wall that has infinite mass 
(or, better still in this framework, infinite viscosity) and on the last is exerted a
force $f$.
The Langevin equations describing such a polymer can be written as:
\begin{equation}
\begin{cases}
\gamma_m \dot{u}_1&=-2k_m u_1+k_m u_2+\eta_1\label{first}\\
&\;\,\vdots\\
\gamma_m \dot{u}_n&=-2k_m u_n+k_m u_{n-1}+k_m u_{n+1}+\eta_n\\
&\;\,\vdots\\
\gamma_m \dot{u}_N&=-k_m u_N+k_m u_{N-1}+f+\eta_N\, ,
\end{cases}
\end{equation}
where the $\eta_i$ are white Gaussian noises of zero mean and variance
\beq
\la \eta_i(t) \eta_j(0) \ra = 2 k_B T \d_{ij} \d(t) \ .
\eeq
Let us for the moment neglect the noise term. Then, defining $\t_m=\g_m/k_m$,
we can rewrite formally these equations as
\beq
\t_m \dot{u}_n=-2 u_n+ u_{n-1}+ u_{n+1} \ , \hskip1cm \forall n \ ,
\eeq
supplemented by the boundary conditions
\beq\label{boundary}
u_0 \equiv 0 \ , \hskip1cm u_{N+1} \equiv u_N +f/k_m \ .
\eeq

A standard way to find the normal modes of the linear system above is to
search for solutions of the form $u_n(t) = u_n(0) \exp(-\l t/\t_m)$. One can
easily show that the general solution satisfying the first boundary
condition $u_0=0$ has the form
\beq\label{sistemau}
\begin{split}
u_n(t) &\propto \sin(q n) \exp(-\l(q) t/\t_m) \ , \\
\l(q) &= 2 (1-\cos(q)) \ .
\end{split}\eeq
The second boundary condition (\ref{boundary}) requires that 
$u_{N+1}(t)-u_N(t) = f/k_m =$ const. Since we can always add the constant
value to $u_{N+1}(t)$, we can replace this boundary condition by
$u_{N+1}(t)=u_N(t)$. This requires that $\sin(q N) \sim \sin(q (N+1))$,
then $q = (\pi/2 + p \pi)/N$. The slowest mode then correspond to
$q=\pi/2/N$, that for large $N$ gives a relaxation time
\beq\label{tauN}
\t(N) = \t_m/\l(\pi/2/N) \sim \frac{4}{\pi^2} \t_m N^2 \ .
\eeq
which proves the validity of the scaling in Eq.(\ref{tau3}).

\subsubsection{Recurrence equations for a fixed end}

We want now to write a system of coupled equations for a certain number
of points on the polymer by integrating
out the $u$'s we are not interested in.
To begin, we focus on the end point $u_N$.

It is convenient to perform a Laplace transformation and write
\beq\label{Laplace}
u_n(t) = \int_0^\io d\l \, u_n(\l) e^{-\l t/\t_m} \ .
\eeq
Then Eq.(\ref{sistemau}) becomes in Laplace space
\beq\label{sistemaul}
(2-\l) u_n(\l) = u_{n+1}(\l)+u_{n-1}(\l) \ ,
\eeq
with the same boundary conditions $u_0(\l) \equiv 0$, 
and $u_{N+1}(\l)-u_N(\l) = (f/k_m) \d(\l)$. For $\l\neq 0$ the latter
condition reduces to $u_{N+1}(\l) = u_N(\l)$ as discussed
above for the normal mode analysis.

We introduce a function
\beq
\z_{n-1}(\l) = u_{n-1}(\l)/u_n(\l) \ .
\eeq
Substituting the latter relation in (\ref{sistemaul}) we
get
\beq\label{recapp}
(2 - \l - \z_{n-1}(\l))u_n(\l) = u_{n+1}(\l) \ ,
\eeq
from which we get a Riccati recurrence equation
\beq\label{recU}
\begin{cases}
\z_0(\l) = 0 \hskip1cm (\text{due to }u_0=0) \ , \\
\z_n(\l) = \frac{1}{2 - \l - \z_{n-1}(\l)} \ .
\end{cases}
\eeq
This recurrence can be solved and the function $\z_n(\l)$
computed for all $n$.

Since we are interested in the large time limit, we can expand
the function $\z_n(\l)$ for small $\l$; we obtain
\beq\label{zetaexp}
\begin{split}
\z_n(\l) &= \frac{n}{n+1} + \frac{n(1+2n)}{6(1+n)} \l \\
&+ \frac{n (6 + 19 n + 16 n^2 + 4 n^3)}{180 (1 + n)} \l^2 + O(\l^3)
 \ .
\end{split}\eeq
One obtains the effective equation 
for $u_N$ by substituting the expression above in (\ref{recapp})
and setting $n=N$.
Keeping only the linear term in $\l$ and the leading terms in 
$N \gg 1$, we get
\beq
\left( 1 + \frac1N - \l \frac{N}3 \right) u_N(\l) = u_{N+1}(\l) \ .
\eeq
Moving back to time domain, we obtain
\beq
\t_m \frac{N}3 \dot{u}_N = -\frac{1}N u_N + (u_{N+1}-u_N) \ ,
\eeq
which is equivalent, using the boundary condition $u_{N+1}-u_N = f/k_m$, 
to
\beq\label{effectiveuno}
\frac{\g_m N}3 \dot{u}_N = - \frac{k_m}N u_N + f \ .
\eeq
In this way we got an effective equation for the endpoint of the polymer
that is still a linear first order differential equation and takes into
account only the slowest mode of the chain.

There is however an inconvenient: in fact a straightforward computation
shows that the relaxation time obtained from Eq.~(\ref{effectiveuno}) is
$\t(N) = \t_m N^2/3$, that differs by a factor $\pi^2/12$ from the correct 
value given by Eq.~(\ref{tauN}). The origin of this discrepancy is clearly
in the fact that the expansion we made in Eq.(\ref{zetaexp}) is not convergent
at fixed $\l$ for $n\to\io$, as successive terms in the series are of order
$n^{2p-1}\l^p$.

Let us then go back to the computation of the normal modes of the system
within this formalism. The second boundary condition $u_{N+1}(\l) = u_N(\l)$
implies $\z_N(\l) =1$. The normal modes are the solutions of this equation
with respect to $\l$. 
One can show from the exact expression of $\z_N(\l)$ that
\beq\label{limiteN}
\lim_{N\to\io} N [\z_N(\wt q^2/N^2)-1] = -\wt q \, \cot(\wt q)
\equiv \wt \zeta(\wt q) \ .
\eeq
The zeroes of this function are $\wt q = \pi/2 + k \pi$;
therefore the solutions of $\z_N(\l)=1$ tend for large $N$ to
$\l = (\pi/2 + p \pi)^2/N^2$, in agreement with the exact result
of the previous section.
An inspection of Eqs.(\ref{zetaexp}) and (\ref{limiteN}) shows
that the small $\l$ expansion of $\z_N(\l)$ is equivalent to
perform a small $\wt q$ expansion of $\wt \z(\wt q)$ in order
to find its first zero. This indeed yields 
$\wt \z(\wt q) \sim -1 +\wt q^2/3$ that gives $\wt q = \sqrt{3}$
for the first zero that gives back $\t(N) = \t_m N^2/3$.

Then one can check that a higher order expansion in $\l$ (or
equivalently in $\wt q$) produces a more accurate result; indeed
the series of $\wt \z(\wt q)$ converges for $\wt q<\pi$ while
the zero is located in $\wt q=\pi/2$. It is easy to show that
if one truncates the series to order $p$, the
difference between the solution and the true zero 
is exponentially small in $p$.

\subsubsection{Discussion}

The conclusion of this section is that {\it Eq.~(\ref{effectiveuno})
is a correct description of the dynamics of the end of the polymer
in the limit of large $N$ and large times}. While it captures the
correct scaling with $N$ of the relaxation time, the coefficient is
wrong by a factor of $\pi^2/12 \sim 0.82$. Still this is quite
satisfactory for our purposes since the experimental error in the
determination of $\t_m$ is of the same order of magnitude.
Better approximations can be obtained by truncating the expansion
of $\z_N(\l)$ to higher orders in $\l$, therefore obtaining a higher
order differential equation for $u_N(t)$.

In the following, we will derive the coupled equation for many
marked points along the chain, limiting ourselves to the first
order truncation. This produces first-order differential equations
of the Langevin type.

\subsection{Dynamics of two coupled polymers}

\begin{figure}
\includegraphics[width=8cm]{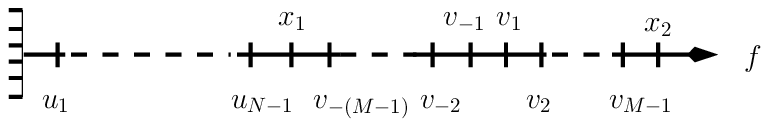}
\caption{Two joint polymers subjected to an external
force $f$. $x_1$ marks the endpoint of the first polymer
made of $N$ links whose endpoints are $u_1,u_2,\cdots,u_{N-1},
u_N \equiv x_1$.
The second polymer originates from $x_1$ and
is made by $2M-1$ links, whose endpoints are 
$v_{-(M-1)},v_{-(M-2)},\cdots,v_{-1},v_1,\cdots,v_{M-2},v_{M-1},x_2$.}
\label{fig:polyintegra}
\end{figure}

We will now show how to use this formalism to derive
coupled equations for different points on a composite polymer.
We keep neglecting the noise, that we will reintroduce at the
end of this section.

As a simple example, let us consider the polymer drawn in figure
\ref{fig:polyintegra}. It is composed by $N$ monomers of 
type ``U''
linked to $2M-1$ monomers of type ``V''. The two types of monomers
might differ for the value of the microscopic sping constant, bead
viscosity etc. If the monomers are identicals, then we are just
marking a point in the middle of a polymer. 

The effective equation for the endpoint of polymer U can be derived
following the analysis of the previous section. We denote $x_1 \equiv
u_N$ and we get
\beq\label{eqUeff}
\g_m^U \frac{N}3 \dot{x_1}(t) = -\frac{k^U_m}N x_1(t) + 
k_m^V (v_{-(M-1)}(t)-x_1(t)) \ ,
\eeq
where the last term is the ``external'' force that the polymer V exerts
on U.

\subsubsection{Integration of the V polymer}

Now we want to integrate out all the monomers $v_{-(M-1)}, \cdots, v_{M-1}$
in order to obtain the coupling between $x_1$ and $x_2$. To this aim,
and in order to keep the formalism symmetric,
we can start from the middle of the polymer V 
by integrating simultaneously $v_{-1}$ and $v_1$ in order to obtain
effective equations for $v_{-2}$ and $v_2$, and so on.
In Laplace space (note that now in Eq.~(\ref{Laplace}) 
$\t_m = \t_m^V$)
the equations for $v_{\pm 1}$ have the form
\beq\label{vuno}
\begin{split}
&(2-\l) v_{-1}(\l) = v_{-2}(\l) + v_1(\l) \ , \\
&(2-\l) v_{1}(\l) = v_{2}(\l) + v_{-1}(\l) \ .
\end{split}\eeq
These can be easily solved to get $v_{\pm 1}$ as function
of $v_{\pm 2}$. Iteration leads to the following form for the
equation after $n$ steps:
\beq\label{viter}
\begin{split}
&\xi_n(\l) \, v_{-n-1}(\l) = v_{-n-2}(\l) + \h_n(\l) \, v_{n+1}(\l) \ , \\
&\xi_n(\l) \, v_{n+1}(\l) = v_{n+2}(\l) + \h_n(\l) \, v_{-n-1}(\l) \ .
\end{split}\eeq
One can check that this form is stable under one step of iteration
and the following recursion relations are obtained:
\beq\label{recV}
\begin{cases}
\xi_0 = 2 - \l \ , \\
\eta_0 = 1 \ , \\
\xi_{n+1} = 2- \l - \frac{\xi_n}{\xi_n^2-\eta_n^2} \ , \\
\eta_{n+1} =  \frac{\h_n}{\xi_n^2-\eta_n^2} \ , \\
\end{cases}
\eeq
where the initial
values are determined by consistency 
between (\ref{vuno}) and (\ref{viter}) for $n=0$.
These recurrences are easily solved by introducing the two quantities
$A_n = 1/(\xi_n-\h_n)$ and $B_n=1/(\xi_n+\h_n)$; these satisfy the same
recurrence in (\ref{recU}) except for the initial condition which is
different and determined according to (\ref{recV}).

At the leading order in $n \to \io$ and at first order in $\l$, we
get
\beq\label{Vfirstl}
\begin{split}
&\xi_n(\l) = 1+\frac{1}{2n} - \frac{2n}3 \l \ , \\
&\eta_n(\l) = \frac{1}{2n} + \frac{2n}6 \l \ .
\end{split}\eeq
Finally one obtains from this procedure a coupled equation for 
$v_{-(M-1)}$ and $v_{M-1}$ where also $x_1\equiv v_{-M}$ and 
$x_2 \equiv v_M$ appear. 

\subsubsection{Coupled effective equations}

To obtain the coupled effective equations one starts
from the following system:
\beq\label{A22}
\begin{cases}
-\g^U_m \frac{N}3 \frac{\l}{\t^V_m} x_1 = -\frac{k^U_m}N x_1 + 
k_m^V (v_{-M+1}-x_1) \ , \\
\xi_{M-2}(\l) \, v_{-M+1}(\l) = x_1 + \h_{M-2}(\l) \, v_{M-1}(\l) \ , \\
\xi_{M-2}(\l) \, v_{M-1}(\l) = x_2 + \h_{M-2}(\l) \, v_{-M+1}(\l) \ ,\\
(1-\l)x_2 = v_{M-1} + f \ , \\
\end{cases}
\eeq
where the first equation is just the Laplace transform of 
Eq.~(\ref{eqUeff}) (recall that we are using the definition
of Laplace transform (\ref{Laplace}) with $\t_m=\t^V_m$),
the second and third equations are Eq.(\ref{viter}) for $n=M-2$
and the last equation is the Laplace transform of the equation for
$x_2$, that in time domain reads $\g_m^V \dot{x_2} = -k^V_m (x_2 -v_{M-1})
+ f$.

Eliminating $v_{-M+1}$ and $v_{M-1}$ from these equations, using the
recurrence equations (\ref{recV}) and the result (\ref{Vfirstl}) we
finally get the coupled equations:
\beq\label{coupled-time}
\begin{cases}
\left(\g^U_m \frac{N}3 + \g^V_m \frac{2M}3 \right) \dot{x_1}
+ \g^V_m \frac{2M}6 \dot{x_2} = - \frac{k^U_m}N x_1 + 
\frac{k^V_m}{2M} (x_2-x_1) \ , \\
\g^V_m \frac{2M}3 \dot{x_2} +  \g^V_m \frac{2M}6 \dot{x_1}
= - \frac{k^V_m}{2M} (x_2-x_1) + f \ .
\end{cases}
\eeq

At this point we reintroduce the free energy of the polymer chain,
defining $N_1 \equiv N$ and $N_2 \equiv 2M-1 \sim 2M$:
\beq
F(x_1,x_2) = \frac{k^U_m}{2N_1} x_1^2 + \frac{k^V_m}{2 N_2} (x_2-x_1)^2 
\ ,
\eeq
and a matrix
\beq\label{gamma22}
\G \equiv 
\begin{pmatrix}
\g^U_m \frac{N_1}3 + \g^V_m \frac{N_2}3 & \g^V_m \frac{N_2}6 \\
\g^V_m \frac{N_2}6 & \g^V_m \frac{N_2}3 \\
\end{pmatrix}
\eeq
so that we can write the system above as
\beq\label{finalchain}
\G_{ij} \dot{x_j} = -\frac{\partial F}{\partial x_i} + f_i + \eta_i \ ,
\eeq
where $\vec{f} = (0,f)$ is the external force vector and we
reintroduced the noise term $\vec{\h}$ that we neglected before.

The correlation function of the noise at this point is determined
by the requirement that the fluctuation-dissipation relation is
verified. This imposes that
\beq\label{correnoise}
\la \h_i(t) \h_j(0) \ra = 2 k_BT \G_{ij} \d(t) \ .
\eeq

\subsection{Beads}

At this point we should add the beads that are used for the optical
manipulation of polymers. These beads are optically tweezed or
subjected to magnetic fields in order to apply forces on the 
polymers. In the former case, the force acting on the bead is a
harmonic force $f = - k (x - X)$, while in the latter
it is constant, $f = f_{ext}$.
Each bead is characterized by a friction coefficient
that can be computed using the Stokes law; 
we denote it by $\g$. Typically they are of the order of
$10^{-5}$ pN s/nm, \ie much bigger than the microscopic viscosity
of the polymers $\g_m \sim 10^{-8}$ pN s/nm.

In presence of a bead attached to the endpoint of a polymer,
the equations of motion (\ref{first}), (\ref{vuno}), etc.
remain valid, but one should add to the coordinate describing
the position of the bead the contribution $\g$ to the viscosity.
For instance, if there is a bead attached to the endpoint $u_N$,
the last Eq.~(\ref{first}) reads
\beq
(\g + \gamma_m) \dot{u}_N=-k_m u_N+k_m u_{N-1}+f+\eta_N \ .
\eeq
Then the above derivation still holds because the last equation
is not used until the end. The only modification will be the
inclusion of $\gamma$ on the diagonal element $\G_{ii}$
corresponding to the coordinate of the bead.

Therefore to describe the beads attached to the end of the
molecular construction in figure~\ref{fig:setup}, we modify
the matrix $\G$ as above; and in case A, we add to the free
energy a term $\frac{1}{2} k (x_4 - X)^2$, while in case
$B$ we add a term $-f_{ext} x_3$. 

In the case of figure~\ref{fig:setup}A, one also has to
include the left bead. In this case, if we call $V$ the
first polymer after the bead, we can start from a system
of equations identical to (\ref{A22}), but with the
first equation replaced by
\beq
-\g \dot{x_1} = -k x_1 + k_m^V (v_{-M+1}-x_1) \ .
\eeq
This will lead again to (\ref{finalchain}) with
\beq
\G \equiv 
\begin{pmatrix}
\g + \g^V_m \frac{N_2}3 & \g^V_m \frac{N_2}6 \\
\g^V_m \frac{N_2}6 & \g^V_m \frac{N_2}3 \\
\end{pmatrix}
\eeq
and
\beq
F(x_1,x_2) = \frac{k}{2} x_1^2 + \frac{k^V_m}{2 N_2} (x_2-x_1)^2 
\ ,
\eeq

\subsection{Description of a generic setup}

The arguments of the previous section suggest that in the general 
case a bead can be treated ``as a particular instance of a polymer''.
In other words, we can consider the setups in figure~\ref{fig:setup}
as chains of $p$ joint elements $U = U_1, U_2, \cdots, U_p$;
each element can be an ``optical trap'' (\ie a spring) or
a polymer of $N_1, N_2, \cdots, N_p$ monomers respectively
(in the case of an optical trap we set by default $N_i =1$).
The endpoint of each element is denoted by $x_i$ and
$\vec{x}\equiv (x_1, x_2, \cdots, x_p)$ is the state vector of the
system
(we also define $x_0 \equiv 0$).

The total free energy is then
$F(\vec{x}) = \sum_{i=1}^p W_{U_i}(x_i-x_{i-1})$
where $W_{U_i}(x) = \frac12 k x^2$ for an optical trap of stiffness
$k$. Then Eq.~(\ref{finalchain})
holds, with $i,j$ running from $1$ to $p$ and the noise correlation
matrix is given by (\ref{correnoise}).

The matrix $\G$ must be constructed as follows. Each diagonal
term $\G_{ii}$, related to $x_i$, is the sum of a Stokes
term coming from a bead possibly attached to $x_i$,
and the contribution 
coming from the two elements adjacent to $x_i$ (except for $i=p$
when there is only one contribution):
\beq\label{gammadiag}
\G_{ii} = \g + \g_m^{U_i} \frac{N_i}{3} + 
\g_m^{U_{i+1}} \frac{N_{i+1}}{3} (1-\d_{ip}) \ ;
\eeq
(the first term is present only if there is a bead attached to $x_i$).
All the off-diagonal elements are zero except the ones adjacent
to the diagonal (i.e. connecting $x_i$ and $x_{i\pm 1}$) 
which get a contribution from the polymer connecting these two ends:
\beq\label{gammandiag}
\G_{i,i+1} = \G_{i+1,i} = \g^{U_{i+1}}_m \frac{N_{i+1}}6 \ ,
\hskip.5cm i = 1,\cdots,p-1 \ . 
\eeq
Note that this final formulation is independent of the Gaussian
form of $F(\vec{x})$ that we assumed in the derivation, therefore
we will use it also for non-Gaussian polymers substituting the
appropriate form of $F(\vec{x})$ in Eq.~(\ref{finalchain}).

To conclude this section, note that a further check of the quality 
of the first order approximation can be done as follows. 
If we consider a single polymer
made of $N_1+N_2$ bases, the corresponding relaxation time is predicted
to be $\t=\t_m (N_1+N_2)^2/3$. On the other hand, we could consider two
coupled polymers of $N_1$ and $N_2$ bases following 
Eq.~(\ref{coupled-time}) for $y^{U,V}_m=\g_m$ and $k^{U,V}_m=k_m$.
The coupled equation can be exactly solved and yield two distinct 
relaxation times (that typically differ by a factor of 10); the
slowest relaxation time can be compared with $\t=\t_m (N_1+N_2)^2/3$.
We found that the difference is at most $20\%$, and the error is
maximal for $N_1 \sim N_2$ while it decreases when one of the
two polymers is much longer than the other.

\section{Transition rates for the fork dynamics}
\label{app:bd}

We now consider a fork $n$ attached to the polymers.
For simplicty we consider the case of a single polymer
whose extension is $x$ and free energy is $W(x,n)$.
We want to construct a stochastic process that samples the equilibrium distribution
$P_{eq}(x,n)=e^{-\beta W(x,n)-G(n;B)}/Z$, where $-G(n;B)$ is 
the free energy gain in closing the first $n$ bases of
DNA, as defined in Eq.~(\ref{p}).

The random process is constructed as follows. The Langevin equation discussed in the 
previous section is discretized with time step $\Delta t$.
If at a given time $t$ the system is in a state $(x,n)$, we allow three possible transitions:
\begin{itemize}
\item $(x,n) \to (x+\D x,n)$ with rate $H^{s}(x,n,\D x)$, 
\item $(x,n) \to (x+\D x,n+1)$ with rate $H^{o}(x,n,\D x)$, 
\item $(x,n) \to (x+\D x,n-1)$ with rate $H^{c}(x,n,\D x)$.
\end{itemize}
We must have
\beq
\int d\D x H^{s}(x,n,\D x)+H^{o}(x,n,\D x)+H^{c}(x,n,\D x) = 1
\eeq

Moreover we can define rates $r^{s,o,c}(x,n) = \int d\D x H^{s,o,c}(x,n,\D x)$ that represent
the rates to stay, open or close $n$ independently of $\D x$. In a practical implementation
we first decide whether to open, close or stay according to $r^{s,o,c}$, and then extract
$\D x$ from the distribution $H^{s,o,c}(x,n,\D x)/r^{s,o,c}(x,n)$.

The detailed balance conditions read
\beq\begin{split}
 P(n,x) H^o(x,n,\D x) &=\\ P(n+1,&x+\D x) H^c(n+1,x+\D x ,-\D x) \\
 P(n,x) H^c(x,n,\D x) &=\\ P(n-1,&x+\D x) H^o(n-1,x+\D x,-\D x) \\
 P(n,x) H^s(x,n,\D x) &=\\ P(n,x&+\D x) H^s(n,x+\D x,-\D x) \\
\end{split}\eeq

We assume that the rate for opening is given by the product of a term
that only depends on the binding free energy as in Eq.~(\ref{ratemd}) and a
term corresponding to a standard Langevin step:
\beq\begin{split}
H^o&(x,n,\D x) = r \D t \, e^{G(n;B)-G(n+1;B)} \, \\ &\times \sqrt{\frac{4 \pi T \D t}{\g_n}} 
\exp\left[ -\frac{\g_n}{4 T \D t} \left( \D x - \frac{f(x,n) \D t}{\g_n}  \right)^2 \right]
\end{split}\eeq
Note that integrating over $\D x$ we find $r^o(x,n)=r \D t \, e^{G(n;B)-G(n+1;B)} = 
r \D t \, e^{-g_0(b_{n+1},b_{n+2})}$, 
consistently with Eq.~(\ref{ratemd}).

Now it is easy to show that the following expression for $H^c(x,n,\D x)$ follows from the second
detailed balance condition:
\beq\begin{split}
H^c&(x,n,\D x) = r \D t \, e^{\b W(x,n)-\b W(x+\D x,n-1)} \, \\ &\times \sqrt{\frac{4 \pi T \D t}{\g_{n-1}}} 
\exp\left[ -\frac{\g_{n-1}}{4 T \D t} \left( \D x + \frac{f(x+\D x,n-1) \D t}{\g_{n-1}}  \right)^2 \right]
\end{split}\eeq
and that the first condition is then automatically satisfied.
Up to now we did not specify the form for $f(x,n)$.
However for a generic $f(x,n)$ the rate above is not Gaussian. 
To obtain a Gaussian rate we assume that
\beq\label{forza}
f(x,n)=-\frac{\partial W(x,n)}{\partial x} \ ,
\eeq
and perform the following simplifications assuming that $\D t$ is small:
\beq\begin{split}
&H^c(x,n,\D x) = r \D t \, e^{\b W(x,n)-\b W(x,n-1)} \\ &\times e^{\b W(x,n-1)-\b W(x+\D x,n-1) - \b f(n-1,x+\D x)} \\ \, 
&\times \sqrt{\frac{4 \pi T \D t}{\g_{n-1}}} 
\exp\left[ -\frac{\g_{n-1}}{4 T \D t} \left( \D x - \frac{f(x+\D x,n-1) \D t}{\g_{n-1}}  \right)^2 \right] \\
\sim& r \D t \, e^{\b W(x,n)-\b W(x,n-1)} \\
&\times\sqrt{\frac{4 \pi T \D t}{\g_{n-1}}}
\exp\left[ -\frac{\g_{n-1}}{4 T \D t} \left( \D x - \frac{f(x+\D x,n-1) \D t}{\g_{n-1}}  \right)^2 \right. \\ &+ \left.
\frac{\b}2 \frac{\partial^2 W(x,n-1)}{\partial x^2} \D x^2\right] \\
\end{split}\eeq
Neglecting $O(\D x^3)$ one obtains a Gaussian distribution for $\D x$, and computing first and second
moment of the Gaussian one can see that at the lowest order in $\D t$ it is equivalent to
\beq\begin{split}
H^c&(x,n,\D x) = r \D t \, e^{\b W(x,n)-\b W(x,n-1)} \, \\ & \times
\sqrt{\frac{4 \pi T \D t}{\g_{n-1}}}
\exp\left[ -\frac{\g_{n-1}}{4 T \D t} \left( \D x - \frac{f(x,n-1) \D t}{\g_{n-1}}  \right)^2 \right]
\end{split}\eeq
From the above expression we deduce that the rate for closing is $r^c(x,n)= r \D t \, e^{\b W(x,n)-\b W(x,n-1)}$;
and one has {\it first} to close, and then perform a Langevin step with force $f(x,n-1)$ and friction $\g_{n-1}$.

Finally, the rate at constant $n$ is simply given by
\beq\begin{split}
H^s&(x,n,\D x) = [1-r^o(x,n)-r^c(x,n)] \, \\ & \times  \sqrt{\frac{4 \pi T \D t}{\g_{n}}}
\exp\left[ -\frac{\g_{n}}{4 T \D t} \left( \D x - \frac{f(x,n) \D t}{\g_{n}}  \right)^2 \right] \ ,
\end{split}\eeq
and it is easy to see that this verifies the third detailed balance equation if Eq.(\ref{forza}) holds and
higher orders in $\D t$ are neglected.

To resume, the implementation of the algorithm is the following:
\begin{enumerate}
\item Choose if stay, open or close, with rates $r^{s,o,c}(x,n)$ respectively.
\item If open, {\it first} perform a Langevin step at $n$ and {\it then} increase $n$ by one.
\item If close, {\it first} decrease $n$ by one and {\it then} perform a Langevin step at $n-1$.
\item If stay, just perform a Langevin step at $n$.
\item Goto 1.
\end{enumerate}

The extension of the above derivation to a case where many polymers are present
is straightforward, since the only polymers whose rate are coupled with $n$ are
the two adjacent ones. All the other polymers are not influenced by $n$ and one can
use standard discretized Langevin dynamics.

\bibliographystyle{unsrt} 
\bibliography{bibliodna}

\end{document}